\documentclass[aps,pra,superscriptaddress,floatfix,twocolumn,notitlepage,longbibliography]{revtex4-1}

\usepackage{graphicx,graphics,epsfig,subfigure,times,bm,bbm,amssymb,amsmath,amsfonts,mathrsfs}
\usepackage[scr=boondoxo,scrscaled=1.05]{mathalfa}
\usepackage[matrix,frame,arrow]{xypic}
\usepackage[pdfstartview=FitH]{hyperref}
\usepackage[pdftex]{color}
\usepackage{dsfont}
\usepackage[export]{adjustbox}
\usepackage[table]{xcolor}

\newcommand{\beq}{\begin{equation}}
\newcommand{\eneq}{\end{equation}}
\newcommand{\beqnn}{\begin{equation*}}
\newcommand{\eneqnn}{\end{equation*}}
\newcommand{\beqy}{\begin{eqnarray}}
\newcommand{\eneqy}{\end{eqnarray}}
\newcommand{\beqynn}{\begin{eqnarray*}}
\newcommand{\eneqynn}{\end{eqnarray*}}

\newcommand{\ket}[1]{ | #1 \rangle  }
\newcommand{\bra}[1]{ \langle  #1 | }

\newcommand{\erf}[1]{Eq. (\ref{#1})}

\def\XXint#1#2#3{{\setbox0=\hbox{$#1{#2#3}{\int}$}
     \vcenter{\hbox{$#2#3$}}\kern-.5\wd0}}

\newcommand{\bes} {\begin{subequations}}
\newcommand{\ees} {\end{subequations}}
	\newcommand{\bea} {\begin{eqnarray}}
	\newcommand{\eea} {\end{eqnarray}}

\newcommand{\red}[1]{\textcolor{red}{#1}}

\newcommand{\ignore}[1]{}

\begin{document}

\title{Extending comb-based spectral estimation to multiaxis quantum noise}

\author{Gerardo A. Paz-Silva}\thanks{These authors contributed equally to this manuscript.}
\affiliation{{Centre for Quantum Dynamics \& 
Centre for Quantum Computation and Communication Technology,  
Griffith University, Brisbane, Queensland 4111, Australia}} 

\author{Leigh M. Norris}\thanks{These authors contributed equally to this manuscript.}
\affiliation{\mbox{Department of Physics and Astronomy, Dartmouth College, 6127 Wilder Laboratory, 
Hanover, New Hampshire 03755, USA}}

\author{F\'elix Beaudoin}
\affiliation{\mbox{Department of Physics and Astronomy, Dartmouth College, 6127 Wilder Laboratory, 
Hanover, New Hampshire 03755, USA}}
\affiliation{NanoAcademic Technologies, 666 rue Sherbrooke Ouest, Suite 802, Montr\'eal, Qu\'ebec, Canada H3A 1E7}

\author{Lorenza Viola}
\affiliation{\mbox{Department of Physics and Astronomy, Dartmouth College, 6127 Wilder Laboratory, 
Hanover, New Hampshire 03755, USA}}

\begin{abstract} We show how to achieve full spectral characterization of general multiaxis additive noise. Our pulsed spectral estimation technique is based on sequence repetition and frequency-comb sampling and is applicable even to models where a large qubit energy-splitting is present (as is typically the case for spin qubits in semiconductors, for example), as long as the noise is stationary and a second-order (Gaussian) approximation to the controlled reduced dynamics is viable. Our new result is crucial to extending the applicability of these protocols, now standard in dephasing-dominated platforms such as silicon-based qubits, to experimental platforms where both $T_1$ and $T_2$ processes are significant, such as superconducting qubits.   
\end{abstract}

\date{\today}
\maketitle

\section{Introduction}

Improving the coherence properties of quantum systems in the presence of unwanted noise is a key step toward realizing the full potential of quantum technologies~\cite{dowling2003quantum}. In particular, obtaining a quantitatively accurate characterization of noise is instrumental to validate theoretical modeling and prediction as well as to design physical-layer quantum control strategies that are optimally tailored to realistic time-dependent noise environments. Acquiring this knowledge is the overarching goal of {\it quantum noise spectroscopy} (QNS), a body of techniques through which noise spectra or correlation functions are estimated based on measurements of dynamical observables of the quantum system of interest (a single qubit sensor in the simplest case) under appropriately chosen external controls and measurements~\cite{Bylander2011,Alvarez2011,Almog2011,Spectro11,Norris2016,Ferrie2018,Liu2019}. In conjunction with algorithmic error mitigation that can be achieved through proper quantum-circuit design and compiling \cite{ErrorMitigation,Endo2018}, spectral properties inferred from QNS are expected to play an important role in enabling near-term intermediate-scale quantum information processors \cite{NISQ}. Ultimately, directly probing the behavior of noise correlations through multiqubit QNS may prove crucial in determining the viability of large-scale fault-tolerant quantum computation \cite{Preskill2013} and in overcoming the effect of correlated quantum noise in entanglement-assisted metrology \cite{Felix2018}.

In recent years, a large focus of QNS has been on characterizing {\em dephasing noise}, that is, noise that couples exclusively along the system's quantization axis and results in transverse relaxation (``$T_2$ effects'') -- under the assumption that any additional, off-axis noise source is either a priori negligible or the resulting longitudinal relaxation (``$T_1$ effects'') can be parametrically incorporated within an ad hoc,  phenomenological model. Though more general noise models have been characterized in contexts where off-axis contributions are tunable \cite{Bylander2011} or using full quantum process tomography \cite{TTM}, most QNS protocols to date involve \emph{single-axis spectral estimation.} These methods have been developed and implemented following two main paradigms: multipulse approaches inspired by dynamical decoupling (DD), in which control sequences consisting of nearly-instantaneous pulses are applied~\cite{yuge2011measurement,Bylander2011,Alvarez2011,QNSReview}; and continuous-wave (CW) approaches inspired by ``spin-locking'' relaxometry~\cite{Yan2013,yan2018distinguishing,willick2018efficient}, in which the control is, typically, a resonant radio-frequency or microwave drive with constant amplitude and noise is probed during driven evolution. In both cases, the basic idea is to shape the control modulation so that the frequency response of the driven qubit sensor is altered in a desired way~\cite{biercuk2011dynamical}. To date, experimental application of QNS has enabled successful reconstructions of dephasing noise spectra in physical settings as diverse as nuclear magnetic resonance~\cite{Alvarez2011,willick2018efficient}, superconducting qubits~\cite{Bylander2011,Quintana2017}, spin qubits in semiconductors~\cite{Muhonen2014,yoneda2018quantum,chan2018assessment}, trapped ions~\cite{Kotler,Kim2017}, and NV centers in diamond \cite{bar2012suppression,hernandez-gomez2018noise}. QNS protocols for high-order dephasing spectra resulting from non-Gaussian statistics have also been validated experimentally, using engineered noise on a superconducting qubit sensor operated outside of a linear-response regime \cite{sung2019non}.

The assumption of single-axis noise processes is too restrictive from both a conceptual and a practical standpoint, however. Many superconducting qubits, for instance, operate in a regime far from pure dephasing, in which the $T_1$ and $T_2$ time scales due to natural noise processes may be of the same order of magnitude \cite{schreier2008suppressing,barends2013coherent,yan2016flux}. In addition, recent advances with spin qubits in semiconductors \cite{mi2018coherent,landig2018coherent} have relied on hybridizing spin and charge degrees of freedom by exploiting exchange-based interactions or the inhomogeneous magnetic field from a micro-magnet, thereby exposing the logical qubit to both longitudinal and transverse noise due to electric fluctuations or coupling to phonons~\cite{kha2015do,russ2016coupling,mi2018coherent,beaudoin2016coupling}. Since noise along distinct qubit axes may a priori arise from the same physical source (e.g., two-level fluctuators or phonons), the corresponding noise processes need not be uncorrelated. Therefore, a complete characterization of noise processes does not only require the {\em simultaneous reconstruction} of the noise spectra along {\em all} relevant qubit axes, but also mandates the estimation of cross-correlation spectra \emph{between} different noise axes.

In this work, we tackle the problem of {\em multiaxis spectral estimation}, by allowing a single qubit sensor to be exposed to general quantum noise with arbitrary correlations along all three axes of the Bloch sphere. Specifically, in the context of DD-based QNS, we focus on non-parametric spectral estimation by \emph{frequency-comb} techniques, as introduced by Alvarez and Suter \cite{Alvarez2011}. The basic idea, in the simplest setting of a stationary Gaussian dephasing noise process described by a power spectral density (PSD) $S(\omega)$, is that repetition of a sufficiently large number $M\gg1$ of a ``base'' pulse sequence with duration $T_c$ effectively shapes the filter function (FF) that describes the control modulation in frequency space into a frequency comb with narrow teeth. This enables $S(\omega)$ to be sampled over a frequency grid determined by the harmonics, $\omega_ m\equiv m (2\pi/T_c)$, with $m\in {\mathbb Z}$. Aside from its conceptual appeal, comb-based QNS affords a number of advantages. Unlike CW approaches, no weak-coupling approximations are needed, for either a single or multiple qubits, as long as the noise is dephasing and Gaussian, which has enabled theoretical extensions to multiqubit settings -- including access to non-classical (asymmetric) spectra and spatio-temporal harmonic features \cite{Cywinski2,Paz2017,krzywda2018dynamical}. The fact that sequence repetition also enforces the emergence of a comb structure in all FFs relevant to high-order dephasing spectra offers a direct means to probe non-Gaussian classical as well as quantum bosonic environments \cite{Paz2014,Norris2016}. Lastly, compared to other DD-based techniques,  
such as $N$-pulse Carr-Purcell-Meiboom-Gill spectroscopy \cite{Bylander2011}, the frequency-comb approach is less susceptible to spectral leakage \cite{Spectro11}, since it takes higher-order harmonics into account in principle.

In practice, comb-based QNS approaches also face limitations in the presence of realistic timing constraints and non-idealities in control and measurements. For instance, deviations from the ideal frequency comb approximation due to finite number of repetitions may bias or significantly complicate the spectral reconstruction procedure, although some compensation may be possible at the cost of additional measurements and analysis \cite{szankowski2018accuracy}. Most importantly for the present discussion, since the resolution to which the PSD may be sampled is determined by the duration $T_c$ of the base sequence, increasing $T_c$ is the only way to obtain a finer sampling grid, with no additional ``knob'' available to independently tune the maximum range of the reconstruction \cite{Spectro11}. This may lead to a large inversion problem, making the approach vulnerable to ill-conditioning and to the numerical errors that follow. Even if numerical stability may be improved by employing suitable regularization \cite{sung2019non}, use of long evolution times for increased sampling resolution may be incompatible with a pure-dephasing approximation for many systems of interest.

With the above considerations in mind, our main objective here is to determine whether and how comb-based QNS may be extended to the characterization of stationary multiaxis single-qubit noise and, if so, to further understand applicability and limitations in realistic scenarios. To answer these questions, we find it useful to contrast a ``driftless'' setting -- in which the qubit energy splitting $\Omega=0$ and thus no internal qubit dynamics is present --  to one where the qubit energy splitting $\Omega\neq 0$ and cannot be neglected. Even if we assume noise to be Gaussian, both situations require perturbative methods, unlike for comb-based QNS in a pure-dephasing setting. Nonetheless, assuming the time-ordered (Dyson) cumulant expansion that determines expectation values of time-evolved qubit observables may indeed be truncated to the second (Gaussian) order, we find that in the driftless setting, complete reconstruction of \emph{all} multiaxis spectra is possible -- including both classical and quantum spectra, the latter arising from non-commuting noise operators. For arbitrary non-zero qubit energy splitting, all of the spectra can still be reconstructed, in principle, by imposing a {\em synchronization constraint} between the internal and the control dynamics, namely, $\Omega T_c=2\pi k$, with $k\in {\mathbb Z}$. While such a constraint can be hard to meet in practice for realistically large values of $\Omega$, we show that in precisely this case drastic simplifications occur, provided that $\Omega T \gg 1$, with $T=M T_c$ being the total evolution time. Indeed, in this regime,  the contribution of most of the multiaxis spectra becomes negligible, and the reduced qubit dynamics is effectively {\em characterized by only three spectra}: one dephasing spectrum, determined by the two-point autocorrelation of noise operators along the quantization axis (say, $z$), and two generally complex spectra, that result from two-point correlators of off-axis noise (along $x,y$). All non-vanishing spectra can be reconstructed using our multiaxis comb-based approach, in principle.

The content is organized as follows. In Sec.~\ref{secSystemControl}, we describe the relevant open-system model Hamiltonian for the  driven single-qubit sensor. In particular, in addition to the standard representation of multiaxis additive noise in terms of Cartesian components, we introduce a {\it spherical representation} (Sec.~\ref{secInitialConsiderations}), which will be expedient for analyzing QNS in the distinctive dynamical regimes mentioned above. We also describe the available pulse control capabilities (Sec.~\ref{secControlResources}), and give the solution for the reduced qubit dynamics within a cumulant expansion truncated to the second order (Sec.~\ref{secReducedDynamics}). In Sec.~\ref{secTools}, we introduce the noise spectra and the control FFs that are needed for formulating the multiaxis estimation problem in the frequency domain (Sec.~\ref{secFrequencyDomain}), and specify the main steps involved in the frequency-comb QNS approach along with the required control symmetries (Sec.~\ref{basicspectro}). Notably, by borrowing from control techniques for decoupled systems \cite{UniDD}, we also introduce {\it frame-tilting control sequences} (Sec.~\ref{secFrameTilting}), which will be instrumental to achieve the level of flexibility needed for simultaneous multi-spectral reconstruction. In Sec.~\ref{secProtocols}, we provide expressions for the experimentally accessible physical quantities in terms of noise cumulants (Sec.~\ref{secAccessibleQuantities}), and show how the distinction between two types of FFs, which we term {\it balanced} and {\it imbalanced}, is key for devising QNS protocols in the two distinctive energy-splitting regimes (Sec.~\ref{secBalancedImbalanced}). In Sec.~\ref{Illus}, we further explain how to combine experimental measurements and build control sequences to grant access to all the target spectra, and provide an illustrative example of a numerical reconstruction (Sec.~\ref{secSampleReconstruction}). Section~\ref{secLimitations} is devoted to a critical assessment of comb-based QNS methods in the presence of noise supported over a wide frequency band, pointing to more stringent practical limitations in multiaxis as opposed to single-axis settings for many realistic qubit devices.  

\section{System and control setting} 
\label{secSystemControl}

\subsection{Open-system model Hamiltonian}  
\label{secInitialConsiderations}

We consider a  single qubit interacting with an arbitrary environment (or ``bath'') in the presence of open-loop control. In the laboratory frame, the evolution in the joint Hilbert space $\mathcal{H}_S \otimes \mathcal{H}_B$ is ruled by a total Hamiltonian of the form $$H^{\rm{lab}}_{\rm{tot}} (t) = H_S + H_B+ H_{SB}(t) + H_{\rm ctrl}^{\rm lab}(t),$$ where $H_S$ $(H_B)$ dictates the internal system (bath) free dynamics, $H_{SB}(t)$ describes the system-bath interaction, and the control Hamiltonian $H_{\rm ctrl}^{\rm lab}(t)$ acts non-trivially only on $\mathcal{H}_S$. We will work in the basis in which the qubit Hamiltonian is diagonal, and associate the quantization direction with the $\sigma_z$ Pauli matrix.  In this way, in units $\hbar=1$, the internal system dynamics and the system-bath interaction are described, respectively, by Hamiltonians
\begin{equation}
H_S= \frac{\Omega}{2} \sigma_z, \quad H_{SB} (t) =\!\!\sum\limits_{\alpha=x,y,z} \!\sigma_\alpha \otimes B^0_{\alpha}(t),
\label{labH}
\end{equation}
where $\Omega$ is the (known) qubit energy splitting, and $B^0_\alpha(t)= B^{0\,\dag}_\alpha(t)$, for all $t$, are time-dependent bath operators. Formally, we allow for the bath operators to have both a quantum (non-commuting) and a classical (c-number) component, namely, $B_\alpha^0(t) \equiv \widetilde{B}_\alpha^0 (t)+ \zeta_\alpha (t) I_B$, with $I_B$ being the bath identity operator and $\zeta_\alpha(t)$ denoting a classical stochastic process. Thus, the limit of purely classical noise corresponds to $B_\alpha^0(t) = \zeta_\alpha(t) I_B$. Note also that in writing Eq. \eqref{labH}, we assume that Tr$_S(H_{SB}(t))=0$, that is, a purely rank-2 coupling \cite{Paz2017}, \footnote{We remark that the separation into $H_S$ and $H_{SB}$ in Eq. \eqref{labH} is dictated by the problem and task at hand. If the external control Hamiltonian supplies full qubit controllability, one can formally redefine $H_{SB}$ to also include $H_S$. On the other hand, in several physical systems of interest, $H_S$ is necessary to achieve full qubit controllability and cannot be included in $H_{SB}$.}.

Physically, the bath operators $B^0_\alpha(t)$ are responsible for introducing noise effects in the reduced qubit dynamics, in a way that will be made quantitatively precise in Sec. \ref{secReducedDynamics}.  We remark that the definition of what constitutes a transverse (dephasing) vs. a longitudinal (relaxation) decoherence process is tied to the choice of the working frame. Relative to the qubit ($z$) eigenbasis, environmental modes that couple ``on-axis'' in an energy-conserving fashion contribute only to the the transverse relaxation time, $T_2$, whereas ``off-axis'' couplings along $x,y$ are responsible for both an energy-non-conserving contribution to the latter, and a finite longitudinal relaxation time, $T_1$. If the quantization axis is changed, however, as in spin-locking QNS \cite{Yan2013}, an originally purely transverse contribution may induce both dephasing and relaxation in the new frame. 

In the interaction picture ($I$) with respect to the total internal Hamiltonian $H_S+H_B$, the qubit-bath
 dynamics are generated by
\begin{equation}
H_I(t) = \!\!\sum_{\alpha=x,y,z}  \!\!\sigma_\alpha(t) \otimes B_\alpha (t) +H_{\rm ctrl}(t),\notag
\end{equation}
where
\begin{align*}
B_\alpha(t) & \equiv e^{iH_B t} B^0_\alpha(t) e^{-i H_B t}\\
H_{\rm{ctrl}}(t) & \equiv e^{i \Omega t \sigma_z/2}H_{\rm{ctrl}}^{\rm lab}(t)e^{-i \Omega t \sigma_z /2}
\\\sigma_\alpha(t)&\equiv e^{i \Omega t \sigma_z/2}\sigma_\alpha
e^{-i \Omega t \sigma_z/2}.
\end{align*}
In the ``driftless" setting when $\Omega=0$, the interaction Hamiltonian reduces to
\begin{equation}
H_I^{(\Omega=0)}(t) = \!\!\sum_{\alpha=x,y,z}  \!\!\sigma_\alpha \otimes B_\alpha (t) +H_{\rm ctrl}(t).
\label{Hcart}
\end{equation}
When $\Omega\neq 0$, on the other hand, the off-axis couplings induce transitions between the $\pm\Omega$ qubit eigenstates. In this case, it is convenient to re-express the interaction Hamiltonian in terms of ladder operators, 
\begin{equation} 
H_I(t)=\!\!\sum_{j=-1,0,+1} \!\!\sigma_{j} \otimes e^{i j \Omega t}B_{-j}(t) +H_{\rm ctrl}(t).
\label{Hsph}
\end{equation}
Here, we use the standard quantum-mechanical definition of spherical (rank-1) vector-operator components, 
\begin{align*}
v_{\pm 1}(t)\equiv \frac{v_{x}(t)\pm iv_{y}(t)}{\sqrt{2}}, \quad v_{0}(t)\equiv v_{z}(t).
\end{align*}
In what follows, as we set up the QNS problem, we will make reference to both the driftless, Cartesian representation in \erf{Hcart} and the spherical representation in \erf{Hsph}. While doing so may seem excessive at first, we will see later on that different representations facilitate the analysis in different regimes of interest. Accordingly, from here on, we shall use Greek ($\alpha,\beta$) indices when specifically working in Cartesian coordinates, $(j,m)$ to denote spherical coordinates, and Latin ($a,b$) when equations apply to both cases.

\subsection{Control resources}
\label{secControlResources}

In order to more easily formulate and analyze the control problem, as customary we further transform the Hamiltonian in Eq. \eqref{Hsph} (or the simpler version in Eq. \eqref{Hcart}) to the toggling-frame, that is, the interaction picture defined by $H_\text{ctrl}(t)$, which leads to a Hamiltonian of the form
\begin{equation} 
\label{Hami}  H(t) = 
\begin{cases}{} 
\sum\limits_{\alpha,\alpha'=x,y,z} y_{\alpha,\alpha'}(t) \sigma_{\alpha'} B_{\alpha}(t)  & \textrm{(Cartesian),} \\
\sum\limits_{j,j'=-1,0,+1} y_{j,j'}(t)  e^{i j \Omega t} \sigma_{j'} B_{-j}(t) & \textrm{(spherical).} 
\end{cases} 
\end{equation}
Here, following \cite{Paz2014,Paz2017}, the $y_{\alpha,\alpha'}(t)\; (y_{j,j'}(t))$ are switching functions which encapsulate the effect of the applied control in the Cartesian (respectively, spherical) coordinates. Letting  $U_{\rm ctrl}(t) \equiv \mathcal{T} e^{- i \int_{0}^t \mathrm{d}s H_{\rm ctrl}(s)}$ denote the time-ordered unitary control propagator, the switching functions are given by 
$$y_{a,a'}(t) = \frac{1}{2} {\rm tr}[  U^\dagger_{\rm ctrl}(t)  \sigma_a U_{\rm ctrl}(t) \sigma_{a'}^\dagger].$$
Note that the $y_{\alpha,\alpha'}(t)$ are real, whereas the $y_{j,j'}(t)$ are generally complex since $\sigma_{\pm 1}(t)$ are not Hermitian. It is also interesting to note that, for $j=\pm 1$ in Eq. \eqref{Hami}, the corresponding switching functions are effectively multiplied by the time-dependent factor $e^{\pm i \Omega t}$; as we will see, this will have the effect of displacing FFs by $\pm \Omega$ in the frequency domain.

While the formalism we employ to model the qubit dynamics applies to arbitrary open-loop control, the QNS protocol we will present requires a control Hamiltonian capable of generating arbitrary instantaneous qubit rotations, $R(\vec{\theta}) \in \rm{SO}(3)$, at pulse times of our choice. By using the parametrization $R(\vec{\theta}) = R(\theta_x, \theta_y, \theta_z)= R_x(\theta_x) R_y(\theta_y) R_z(\theta_z)$, we denote the corresponding control pulse by  
\begin{equation}
P_{\vec{\theta}} \equiv e^{- i {\sigma_x \theta_x}/{2}} e^{- i {\sigma_y \theta_y}/{2}}e^{- i {\sigma_z \theta_z}/{2}} \in {\rm SU}(2).
\label{Ptheta}
\end{equation}
We point out that this parametrization is a matter of generality and should not be taken to require the execution of three individual pulses; indeed, the sample execution of our protocol to be presented in  Sec.~\ref{Illus} will employ only pulses along a single coordinate axis. The action of pulses $P_{\vec{\theta}}$ on operators in our two bases of interest can be obtained by making use of standard algebraic Pauli identities --  in particular, $\sigma_{\alpha} \sigma_{\alpha'} = \delta_{\alpha,\alpha'} I_S + i \epsilon_{\alpha \alpha' \beta} \sigma_\beta$ and $\sigma_\alpha \sigma_{\alpha'} \sigma_\alpha = \sigma_{\alpha'} ( 2\delta_{\alpha,\alpha'}-1)$. Notable cases are $\pi$-rotations around one of the coordinate axes, e.g., $P_{(0,0,\pi)} \equiv [\pi]_z$. With this notation, $\theta$ rotations around the quantization ($z$) axis induce the transformations
\begin{align*}
P_{(0,0,\theta_z)}=[\theta]_z: \sigma_{\pm} \mapsto  e^{\mp i \theta} \sigma_\pm, \quad   \sigma_0 \mapsto \sigma_0.   
\end{align*}
Importantly, starting from free evolution ($y_{j,'j}(t)=\delta_{j,j'}$), $\pi/2$-pulses allow the generation of purely imaginary-valued switching functions, whilst arbitrary $\theta$ values typically generate complex-valued switching functions. 

Under the above rules,  the toggling frame transformation of Eq. \eqref{Hami} induces particular structures in the switching functions. Specifically, one may verify that:

\smallskip 

(i) In the Cartesian basis, $\pi$-pulses along any coordinate axis $\alpha$, corresponding to $[\pi]_\alpha$, induce digital switching functions $y_{\alpha,\alpha'}(t)$ that take values only in $\{+1,-1\}$ and are ``diagonal'', in the sense that $y_{\alpha,\alpha'}(t) \propto \delta_{\alpha,\alpha'}$. In contrast, non-$\pi$ pulses generally induce induce ``non-diagonal'' switching functions, which can take a continuum of values in $[-1,1]$. 

\smallskip

(ii) In the spherical basis, both $\pi$ and non-$\pi$ pulses induce non-diagonal switching functions $y_{j,j'}(t)$. This stems from the fact that, for example,
\begin{align*}
[\theta]_x :\sigma_{+} & \mapsto \sin^2 \frac{\theta}{2}\, \sigma_{-} +
i \frac{\sin \theta}{\sqrt{2}}\, \sigma_0
+\cos^2 \frac{\theta}{2}\, \sigma_{+},\\
\sigma_0 & \mapsto - i\frac{\sin \theta}{\sqrt{2}}\, \sigma_{-}+
\cos \theta \, \sigma_0
+ i \frac{\sin \theta}{\sqrt{2}}\, \sigma_{+},\\
\sigma_{-} & \mapsto \cos^2 \frac{\theta}{2}\, \sigma_{-} -
i \frac{\sin \theta}{\sqrt{2}}\, \sigma_0
+\sin^2 \frac{\theta}{2}\, \sigma_{+}.
\end{align*}
Notably, switching functions now take values in the set  $\{-1,0,+1\}$ under the action of $\pi$ pulses, while more generally they take on complex values. This will be very important for our filter design stage, as we will see later (Sec.~\ref{secProtocols}). 

In our protocols, we will consider sequences of $p$ pulses of the general form given in Eq. \eqref{Ptheta}, applied at different control times $\{t_i\}$ over a total duration $T$. The interaction-picture unitary propagator associated to such a control sequence then takes the form 
\begin{eqnarray} 
U_I(T) &\!=\!& \!\Big[ P_{\vec{\theta}_{p-1}} U_0(t_{p-2}, t_{p-1}) \cdots P_{\vec{\theta}_1} U_0(t_0,t_1) \Big] 
P_{\vec{\theta}_0} \nonumber \\
& \!\equiv \! &\!\Big[\!\prod_{i={p-1},\ldots,1}  \! P_{\vec{\theta}_i} U_0(t_{i-1}, t_i)\Big] P_{\vec{\theta}_0} \label{IPprop} \\
& = & U_{\rm ctrl}(T)  \,U(T) = \Big[ P_{\vec{\theta}_{p-1}}\!\! \cdots P_{\vec{\theta}_1} P_{\vec{\theta}_0}\Big] U(T). 
 \label{Tprop} 
\end{eqnarray}
Here, we have taken $t_0 =0$, time is understood to increase from right to left, $U_0(t,t')$ in Eq. \eqref{IPprop} denotes free evolution of the system and bath from $t \mapsto t'$ and, in Eq. \eqref{Tprop}, we have explicitly expressed the controlled evolution in terms of the toggling-frame propagator, $U(t) \equiv \mathcal{T} e^{- i \int_{0}^t \mathrm{d}s H(s)}$, with $H(t)$ given in Eq. \eqref{Hami}.

\subsection{Reduced qubit dynamics}  
\label{secReducedDynamics}

Evolution under the natural system Hamiltonian and the control in the presence of the environment induces non-trivial dissipative dynamics on the probe qubit, which is captured by the expectation value of relevant physical observables. Let $O$ denote an invertible qubit observable. Assuming that the system and the bath are initially in a factorized state, $\rho_{SB}(0) \equiv \rho_0 = \rho_S \otimes \rho_B$, the time-dependent expectation value in the toggling frame is determined by 
\begin{equation}
E_{\rho_0} ({O}(t)) = \langle {\rm tr} [ U(t) (\rho_{S} \otimes \rho_B) U(t)^\dagger {O}]\rangle_c, 
\label{expect}
\end{equation}
where $\langle \cdot \rangle_c$ represents an average over realizations of the classical stochastic process $\zeta_b (t)$ and, formally, we let $\rho_B= I_B$ in Eq. \eqref{expect} if noise is purely classical. By also denoting $\langle \cdot \rangle_q \equiv {\rm tr}_B [ \cdot \rho_B]$ and following \cite{Paz2017}, one can rewrite the above expectation value via a cumulant expansion as 
\begin{eqnarray}
\nonumber E_{\rho_0} (O(t)) &= &\langle {\rm tr}_S [\langle O^{-1} U(t)^\dagger {O} U(t) \rangle_q  \rho_{S} O ] \rangle_c \\
&  \equiv & {\rm tr}_S  \bigg[e^{\,\sum\limits_{k=1}^\infty {(-i)^k}\frac{\mathcal{C}_O^{(k)}}{k!} }   \rho_{S} \,{O} \bigg] 
\label{cumulantexp} \\
& = & {\rm tr}_S  \Big[ \langle {O}^{-1} U(t)^\dagger {O} U(t) \rangle \,\rho_S O\Big], \label{compact}
\end{eqnarray}
where in Eq. \eqref{compact} we have introduced the compact notation $\langle \cdot \rangle \equiv \langle\langle \cdot \rangle_c\rangle_q$ to denote taking both averages. 

The $k$-th generalized cumulant $\mathcal{C}^{(k)}$ entering Eq. \eqref{cumulantexp} is an operator on the qubit and may be obtained as follows. Realizing that one can rewrite  ${O}^{-1} U(t)^\dagger {O} U(t)  = \mathcal{T} e^{-i \int_{-T}^T\widetilde{H}_O (s) ds }$ with the operator-dependent effective Hamiltonian 
\begin{equation} 
\widetilde{H}_O(t) = \begin{cases} 
- {O}^{-1}H (T-t) {O} & t \in [0,T],\\
\,\,\,\,\,\,\,\,\,\, \,\,\,\,\,H (T+t) & t \in [-T,0],
\end{cases} \label{effH}
\end{equation}
the cumulant expressions can be obtained from the moment-cumulant generating equation 
\begin{align*}
\langle \mathcal{T} e^{-i \int_{-T}^T\widetilde{H}_O (s) \mathrm{d}s } \rangle &=  e^{\sum\limits_{k=1}^\infty {(-i)^k \frac{\mathcal{C}_O^{(k)}}{k!} }}= \sum_{k=1}^\infty \frac{\mathcal{D}_O^{(k)}}{k!},
\end{align*}
where the terms in the Dyson-like expansion in the right hand-side are given by 
\beq
\label{cumg}
\mathcal{D}_O^{(k)} =  k! \, (-i)^k \int_{-T}^T \mathrm{d}_>\vec{t}_{[k]} \langle \widetilde{H}_O(t_1) \cdots \widetilde{H}_O(t_k)\rangle,
\eneq  
with $\int_{s}^{s'} \mathrm{d}_>\vec{t}_{[k]} \equiv \int_{s}^{s'} \mathrm{d}t_1 \cdots \int_{s}^{t_{k-1}} \mathrm{d}t_k$, and the inclusion of factorial term being dictated by convenience. In general, the expansion in Eq.~\eqref{cumulantexp} involves an infinite number of terms, and can be exactly truncated only in special scenarios. For example, exact truncation to the second order is possible when the noise model is dephasing-only and Gaussian \cite{Paz2017}. Throughout this work, we will not invoke Gaussian noise statistics, but we shall demand that suitable conditions are obeyed to justify truncating the expansion at order two~\footnote{Formally, we are demanding that the system-bath coupling, the controls being used, and the relevant spectra, are such that the (operator) norm 
$ ||\sum  (-i)^k \mathcal{C}^{(k)}/k! ||  \simeq  ||-i \mathcal{C}^{(1)}-{\mathcal{C}^{(2)}}/{2}||$. 
For instance, this will be appropriate in a weak-coupling regime, as discussed e.g. in \cite{Breuer:book}.}.

Under the above assumption, we thus focus only on the first two cumulants. After an appropriate change of variables, allowing us to change the integration domain from $[-T,T]$ to $[0,T]$, and recalling the definition of $\widetilde{H}_O(t)$ in Eq. \eqref{effH}, these can be written as follows:
\begin{eqnarray}
\label{cumulantes1}
\mathcal{C}^{(1)}_O & = & \!\int_{0}^T \!\!\!\mathrm{d}t_1 \Big[C^{(1)}( H(t_1))- C^{(1)}(\overline{H}_O(t_1)) \Big] ,\\
\nonumber \frac{\mathcal{C}_O^{(2)}}{2!} 
& =& \!\int_{0}^T \!\!\! \mathrm{d}_>\vec{t}_{[2]} \, \Big[ C^{(2)}(H(t_1),H(t_2))  \\ 
\label{cumulantes2} & & + \,
C^{(2)}(\overline{H}_O(t_2),\overline{H}_O(t_1))  \\
\nonumber & &
-\, C^{(2)}( \overline{H}_O(t_1), H(t_2)) - C^{(2)}(\overline{H}_O(t_2) , H(t_1))\Big],
\end{eqnarray}
in terms of the toggling frame Hamiltonian $H(t)$ and its observable-conjugated version $\overline{H}_O(t)\equiv {O}^{-1} {H}(t) {O}$, via the cumulant expressions 
\begin{align*}
C^{(1)}(A) &= \langle A \rangle ,\\
C^{(2)}(A,B) & = \langle A B\rangle - \frac{1}{2} \Big(\langle A \rangle \langle B \rangle +\langle B \rangle \langle A \rangle \Big) .
\end{align*}
Notice that the average $\langle \cdot \rangle$ is non-commuting, in the sense that when $A$ and $B$ are operators on $\mathcal{H}_S \otimes \mathcal{H}_B$, one generally has $[ \langle A \rangle,\langle B \rangle] \neq 0$ and care must be taken during calculations.

We remark that, while evaluating qubit time-evolved observables in the toggling frame is mathematically convenient, in experiments these cannot be measured directly.  Nevertheless, they can be related to measurable quantities in either the lab frame or the interaction frame (or, under resonance conditions, a frame which is also co-rotating with the carrier frequency of the control) by additionally implementing an appropriate compensating transformation, leveraging the fact that both the internal qubit Hamiltonian and the applied control propagator in Eq. \eqref{Tprop} are known.  In practice, if sufficiently fast control is available, this amounts to implementing an additional rotation  immediately before measurement that effectively un-does the rotation generated by the ideal control $U_{\rm ctrl}(T)$ \cite{Spectro10}.

\section{Tools for spectral estimation} 
\label{secTools}

\subsection{Frequency domain representation \label{secFrequencyDomain}}

Given the way in which the Hamiltonian enters the reduced dynamics (see Eqs. \eqref{compact} and \eqref{cumulantes2}), one needs to evaluate expressions with the following general structure:
$$\int d_> \vec{t}_{[2]} \langle H(t_1) \overline{H}_O(t_2)\rangle 
= \!\!\!\sum_{a,a',b',b'} \!\! (\sigma_{a'} O^{-1} \sigma_{b'} O )  [I_2]_{a,a',b,b'}, $$
where the integral 
\begin{align*}
[I_2]_{a,a',b,b'} \equiv \int d_> \vec{t}_{[2]} \,\,  y_{a,a'}(t_1)  y_{b,b'} (t_2)  \langle B_{a}(t_1) B_{b}(t_2) \rangle \end{align*}
is common to all the terms in the aforementioned reduced dynamics equations. By moving to the frequency domain, both the bath correlators $\langle B_{b_1}(t_1) \cdots B_{b_k}(t_k)\rangle$ and products of the switching functions $\{y_{a,a'}(t)\}$ wil then enter the dynamics via their corresponding Fourier transform. 

Specifically, the influence of bath correlations on the qubit dynamics is captured by the multi-time Fourier transform
$$\langle B_{b_1} (\omega_1) \cdots B_{b_k}(\omega_k)\rangle = \int_{-\infty}^{\infty} \!\!\mathrm{d} \vec{t} \,\, e^{i \vec{\omega} \cdot \vec{t}} \langle B_{b_1} (t_1) \cdots B_{b_k}(t_k)\rangle.$$
We will assume that our noise is zero-mean and {\em stationary} which, since we are truncating the cumulant expansion at $k=2$, is captured by the relations 
\begin{align*}
\langle B_b (\omega) \rangle & =0, \\
\langle B_{b_1} (\omega) B_{b_2}(\omega')\rangle &= \delta(\omega  + \omega')  S_{b_1,b_2} (\omega),  
\end{align*}
where as usual the relevant PSD is defined as \begin{align*}
S_{b_1,b_2}(\omega)=\int_{-\infty}^\infty \!\! \mathrm{d}\tau\, e^{-i\omega\tau}\langle B_{b_1}(\tau)B_{b_2}(0)\rangle .
\end{align*}
The PSD can be separated into two components, $S_{b_1,b_2}(\omega)=[S^-_{b_1,b_2}(\omega) + S^+_{b_1,b_2}(\omega)]/2$, depending on the commutator and anticommutator of the noise operators,
\begin{align*}
S_{b_1,b_2}^-(\omega)\equiv&\int_{-\infty}^\infty \!\! \mathrm{d}\tau\, e^{-i\omega\tau}\langle [ B_{b_1}(\tau),B_{b_2}(0)]\rangle,\\
    S_{b_1,b_2}^+(\omega)\equiv&\int_{-\infty}^\infty \!\! \mathrm{d}\tau\, e^{-i\omega\tau}\langle\{ B_{b_1}(\tau),B_{b_2}(0)\}\rangle.
\end{align*}
The classical spectrum $S_{b_1,b_2}^+(\omega)$ is present for both quantum and classical noise, while the ``quantum" spectrum $S_{b_1,b_2}^-(\omega)$ vanishes when the noise is classical and, hence, commuting. Notably, $S_{b_1,b_2}^-(\omega)$ is antisymmetric in the frequency domain, a manifestation of its quantum character. We draw a further distinction between spectra whose indices refer to the same noise operator, the ``self"- spectra $S^{\pm}_{b_1,b_2} (\omega) |_{b_2=b_1}$, or different noise operators, the ``cross"- spectra
$S^{\pm}_{b_1,b_2} (\omega) |_{b_2\neq b_1}$ \cite{Paz2017,Felix2018}. The relations between the various spectra depend on the coordinate system: 

\smallskip

(i) In Cartesian coordinates, since $B_\alpha (t)^\dagger = B_\alpha (t)$, we have that $i [B_{\alpha_1}(t), B_{{\alpha_2}}(t')] $ and $\{B_{\alpha_1}(t), B_{{\alpha_2}}(t')\}$ are self-adjoint, which in turn implies that 
$$[S^{\pm}_{{\alpha_1},{\alpha_2}} (\omega)]^*=S^{\pm}_{{\alpha_2},{\alpha_1}} (\omega)
= \pm S^{\pm}_{{\alpha_1},{\alpha_2}} (-\omega). $$ 

\smallskip

(ii) In spherical coordinates, as noted, bath operators corresponding to $j=\pm1$ are not self-adjoint. One has that $B_{j}(t)^\dagger = B_{-j}(t)$, and, consequently, it follows that 
$$[S^{\pm}_{j,l} (\omega)]^* =S^{\pm}_{-l,-j} (\omega) = \pm S^{\pm}_{-j,-l} (-\omega).$$ 

\smallskip

In a similar manner, the effect of the control may be compactly described in the frequency domain in terms of appropriate fundamental FFs~\cite{Paz2014}, 
\begin{align}
\notag F^{(1)}_{a,a'} (\omega,T) &= \!\int_0^T \!\!\mathrm{d}t \,y_{a,a'}(t) e^{i \omega t} ,\\
\label{FFF} F^{(2)}_{a,a';b,b'} (\omega,\omega',T) &= \!\int_0^T  \!\!\mathrm{d}t \int_0^t  \!\!
\mathrm{d}t' \, y_{a,a'}(t) y_{b,b'}(t')e^{i (\omega t + \omega' t')},
\end{align}
with associated first- and second-order FFs given by \cite{multiDD, Paz2017}:
\begin{align}
\label{GFF} G^{-}_{a,a';b,b'} (\omega,\omega',T) & =F^{(2)}_{a,a';b,b'} (\omega,\omega',T) \\ \nonumber & 
 - F^{(2)}_{b,b';a,a'} (\omega',\omega,T) ,\\
 \notag G^{+}_{a,a';b,b'} (\omega,\omega',T) &=F^{(2)}_{a,a';b,b'} (\omega,\omega',T) \\ \nonumber &
 + F^{(2)}_{b,b';a,a'} (\omega',\omega,T)\\
\nonumber &= F^{(1)}_{a,a'} (\omega,T) F^{(1)}_{b,b'} (\omega',T).
\end{align}

Combining our observations in the frequency domain representation, one can calculate the leading cumulants  in Eq.~\eqref{cumulantexp}. First, recalling that the noise is zero mean, one has that $\mathcal{C}^{(1)}_O = 0$ in both coordinate representations, for arbitrary $O$. As to the second cumulant, while general closed-form expressions may be obtained, we provide them for the simpler and practically relevant case of Pauli observables. In spherical coordinates, direct calculation yields 
\begin{align}
\nonumber 
\frac{\mathcal{C}_{\sigma_\gamma}^{(2)}}{2} &=  \sum_{l,l',j,{j'}} \int_{-\infty}^{\infty} \frac{ \mathrm{d}\omega}{2\pi}F^{(2)}_{j,{j'};{l},l'} (\omega+ j \Omega, -\omega + {l} \Omega,T)   \times  \\
& \nonumber\Big[  (\sigma_{j'} \sigma_{l'} - \sigma_\gamma \sigma_{j'} \sigma_\gamma \sigma_{l'})  S_{-j,-{l}} (-\omega)  \\ 
& \label{cumussphe}
+ (\sigma_\gamma \sigma_{l'} \sigma_{{j'}} \sigma_\gamma - \sigma_\gamma \sigma_{l'} \sigma_\gamma \sigma_{{j'}})   S_{-{l},-j} (\omega) \Big],
\end{align}
where the $\Omega$-displacement in the frequency arguments of the fundamental FFs arises due to the $e^{i \Omega t}$ factors in the spherical Hamiltonian [Eq.~\eqref{Hami}]. In cartesian coordinates, we have
\begin{align}
\label{c2carte}
\frac{\mathcal{C}_{\sigma_\gamma}^{(2)}}{2} &= \sum_{\alpha,\beta,\alpha',\beta'}
\sigma_\alpha\sigma_\beta
\int_{-\infty}^\infty \frac{ \mathrm{d} \omega}{4\pi}  \,S^{ f_{\alpha}^\gamma f_{\beta}^\gamma f^{\alpha}_{\beta}}_{\alpha',\beta'} (\omega) \times \\
\nonumber &\Big[ G^{f_{\alpha}^\gamma f_{\beta}^\gamma}_{\alpha,\alpha';\beta,\beta'} (\omega,-\omega,T) + f_{\alpha}^\gamma G^{+}_{\alpha,\alpha';\beta,\beta'} (\omega,-\omega,T) \Big] ,
\end{align}
with $f^\gamma_\alpha = {\rm tr} [ \sigma_\alpha\sigma_\gamma\sigma_\alpha\sigma_\gamma]/2$, which is hauntingly similar to 
the expression derived in \cite{Paz2017} for a multiqubit dephasing model, i.e., when all operators are mutually commuting (see in particular Eq. (27) therein). The only seemingly minor difference is the $f^\alpha_{\beta}$ superscript in $S^{ f_{\alpha}^\gamma f_{\beta}^\gamma f^{\alpha}_{\beta}}_{\alpha',\beta'} (\omega)$ which, however, will have significant consequences in terms of the symmetries needed to extract all the noise spectra (see Sec. \ref{secProtocols}).

Again, we stress that here we are truncating the expansion to the second order based on a suitable convergence argument (for instance, weak coupling) and not because we are invoking a Gaussian property of the noise. In fact, notice that even when the Gaussian assumption is in place, the non-commuting,  multiaxis nature of the noise prevents the cumulant series from truncating exactly, unlike  for single-axis noise~\cite{Norris2016,Paz2017}. 

\subsection{QNS via frequency comb}
\label{basicspectro}

The interplay between FFs and power spectra is the key element in the QNS protocols we consider. The aim of these protocols is to estimate the full set of dynamically relevant spectra $\{ S^{\pm}_{b_1,b_2} (\omega)\}$ by studying the response of the probe system to the control while in the presence of the target noise. Mathematically, this entails two key steps: first, isolating the integrals that involve a spectrum of interest by choosing an appropriate set of observables and initial states; next, deconvolving each such integral to obtain an appropriate estimate. 

The first problem can be addressed by preparing eigenstates of the Pauli basis $\sigma_\alpha$, for $\alpha =x,y,z$, measuring in the Pauli basis, and combining the resulting expectation values into experimentally accessible quantities. The preparation and measurement procedure simplifies if the system has known symmetries, as one may appreciate by considering a purely dephasing classical noise model, i.e., one for which $B_a(t) \propto \zeta_a(t) \delta_{a,z} I_B $. Using control that preserves the dephasing character of the Hamiltonian, e.g., $[\pi]_x$ pulses, the expectation value of $\sigma_x$ when the qubit is initialized in $(I_S+\sigma_y)/2= \ket{+}\bra{+}$ is given by
\begin{align*}
\log [E_{\ket{+}\bra{+}} (\sigma_x (t))] = \log[ {\rm tr} (e^{- {\mathcal{C}^{(2)}}/ 2!  }   \ket{+}\bra{+} \sigma_x ) ] \\
\qquad =  {-\frac{1}{\pi} \int_{-\infty}^\infty \!\! \mathrm{d}\omega \,G^{+}_{z,z;z,z} (\omega,-\omega, T)  S^{+}_{z,z} (\omega)}, 
\end{align*}
i.e., a single preparation and measurement setting is enough to isolate the integral containing the relevant noise information.  

The second issue, namely, extracting the noise information once the integral has been isolated, has been the object of many recent studies~\cite{Bylander2011,Alvarez2011,Norris2016, Paz2017,Cywinski2,Spectro10,Spectro11}; in particular, DD QNS based on a frequency-comb approach has been widely employed \cite{QNSReview}. The basic idea behind such an approach is to ensure that each of the integrals that can be isolated can be further {\it discretized and truncated} in a systematic way, 
by use of control. The objective is to guarantee that any of the aforementioned integrals takes the form
\begin{equation}
\label{discrete}
I= \int_{-\infty}^\infty  \mathrm{d}\omega \,G(\omega) S(\omega) \simeq A \sum_{k = 0}^K G ( k \omega_0) S( k \omega_0),
\end{equation}
where both the fundamental frequency $\omega_0$ and the proportionality constant $A$ are determined by the choice of control, and the highest harmonic $K$ is adjusted so that $S(\omega)$ is reasonably small for $\omega\geq K\omega_0$ (see, however, further discussion in Sec. \ref{secLimitations}). 
  
If this can be achieved, then each integral is basically a linear equation involving the unknown quantities, $S( k \omega_0)$, along with known, tunable, control-dependent coefficients $A  G ( k \omega_0)$. Since a given experiment provides access to the value of $I$, it is then possible to generate a system of linear equations by changing the control being used, from which an estimate of the target power spectra can ultimately be inferred.

The  discretization described above may be achieved by a suitable choice of control. To make contact with existing tools, it is useful to note that, given the form of the dynamical equations, there are two kinds of FFs we have to be concerned with, namely, $G^{+}_{a,a';b,b'} (\omega,\omega',T)$ and $G^{-}_{a,a';b,b'} (\omega,\omega',T)$ -- formally very similar to the setting of multiqubit pure dephasing we analyzed in \cite{multiDD,Paz2017}. As shown there, the key to generating a frequency comb is the ability to design switching functions exhibiting one or more of the following symmetries: 
\begin{itemize} 
\item[(i)] $T_c$ periodicity, i.e.,  $y_{a,a'}(t) = y_{a,a'}(t+T_c),$ for $T_c>0$.

\item[(ii)] Displacement (anti-)symmetry at time scale $[0,\tau]$, i.e., for $t \in [0,\tau/2]$,
$ y_{a,a'}(t) = \pm \,y_{a,a'}(t+\tau/2).$

\item[(iii)] Mirror (anti-)symmetry at time scale $[0,\tau]$, i.e., for $t \in [0,\tau/2]$,
$ y_{a,a'}({\tau}/2-t) = \pm \,y_{a,a'}(t+{\tau}/{2}).$
\end{itemize}
Most relevant to this paper, one can show~\cite{Alvarez2011,multiDD} that if a control sequence of duration $T_c$ is repeated $M \gg 1$ times over a total duration $T=M T_c$, a frequency comb is created in which an  arbitrary first-order FF is directly expressible in terms of its single-cycle counterpart. For FFs of the $G^+$-type, the frequency comb takes the form
\begin{align}
&G^{+}_{a,a';b,b'} (\omega,{ -\omega}, M T_c) 
= \frac{\sin^2 (M \frac{\omega T_c}{2})}{\sin^2 (\frac{\omega T_c}{2})}  G^+_{a,a';b,b'} 
(\omega , {-\omega}, T_c) \nonumber \\
&\simeq \frac{2 \pi M}{T_c} \sum_k \delta (\omega - k  \omega_0) G^{+}_{a,a';b,b'} (\omega,-\omega, T_c),
\label{combg}
\end{align}
\noindent where $\omega_0 = {2 \pi}/{T_c}$. For the $G^-$-type FFs, in addition to control repetition ($T_c$-periodicity), one of the switching functions entering the FF in Eqs.~\eqref{FFF} and \eqref{GFF} must be displacement anti-symmetric and the other symmetric. This symmetry condition combined with repetition generates another comb,
\begin{align*}
& G^{-}_{a,a';b,b'} (\omega,-\omega, M T_c) \\
& \simeq  \frac{2 \pi}{T_c}\sum_k (-1)^k\delta (\omega - k  \omega_0) G^{+}_{a,a';b,b'} (\omega,-\omega, {T_c}/{2}).
\end{align*}
Additionally, mirror symmetry in the interval $[0,T_c]$  $([0,T_c/2])$ can be leveraged to control the real or imaginary character of the filters $ G^+_{a,a';b,b'}(\omega, -\omega, T_c) 
$  $(G^+_{a,a';b,b'}(\omega, -\omega, T_c/2))$. By combining the tools described above, we showed how all the spectra relevant to a multiqubit dephasing Gaussian noise model could be reconstructed~\cite{Paz2017}. A caveat of having to deal with both types of filters is their different $M$ scaling. While one can, in principle, work around it by a careful choice of controls~\cite{Paz2017}, this imposes constraints on the types of sequences that can be used, which in turn leads to complications in reconstructing spectra that are filtered by the $G^-$-type FFs.

In contrast to single-qubit dephasing, the complexity of the multiqubit protocol arises from the fact that the system operators in $H(t)$ are spanned by a richer albeit still commuting algebra. In the single-qubit multiaxis setting, the situation is seemingly more complex as the system part of the Hamiltonian is additionally spanned by an algebra which is {\em non-commuting}. Thus, one would expect that, at the very least, all of the same symmetries should be necessary in our current scenario. Surprisingly, as we will see in Sec~\ref{secBalancedImbalanced}, the structure of Eq.~\eqref{c2carte} implies that {\em only} $G^+$ filters are relevant for our purposes, which in turn implies the need for a smaller set of symmetries, namely, repetition and mirror, and thus the $M$-scaling complication we highlighted earlier is absent. 
\subsection{Frame-tilting control sequences \label{secFrameTilting}}

While repetition of base sequences exhibiting the aforementioned symmetries is essential to generate the comb, the choice of pulse types in the base sequences is crucial to access all of the spectra, whenever multiple kinds of spectra simultaneously influence the dynamics. In \cite{Paz2017}, an important design principle was the use of ``non-diagonal'' control sequences, capable of generating non-trivial switching functions $y_{a,b}(t)$ not proportional to $\delta_{a,b}$.

Here, we further introduce more general ``frame-tilting'' control sequences, which contain the aforementioned approach as a special case and are closely related to `
`twisted decouplers''~\cite{UniDD}. Recall from \erf{Tprop} that after the application of $i$ pulses at time $t_i$, the control propagator is given by $U_\text{ctrl}(t_i)=P_{\vec{\theta}_{i-1}}...P_{\vec{\theta}_{0}}$. In a tilted decoupling sequence, the trick is to choose the pulses $P_{\vec{\theta}_{i-1}}, ..., P_{\vec{\theta}_{0}}$ so that at each $t_i$, the control propagator can be written in terms of a static ``tilting pulse" $P_\text{tilt}$ and a desired ``target pulse" $P_{\vec{\theta'_i}}$ as follows,
\begin{align}
U_\text{ctrl}(t_i)=P_{\vec{\theta}_{i-1}}...P_{\vec{\theta}_{0}}=P_\text{tilt}P_{\vec{\theta'_i}}.
\end{align}
At each time $t_i$, the control propagator then induces the toggling frame transformation
$$ \sigma_a \mapsto U_\text{ctrl}^\dag(t_i) \sigma_a U_\text{ctrl}(t_i) =P_{\vec{\theta'_i}}^\dag \left[P_{\rm tilt}^\dag  \sigma_a {P_{\rm tilt}} \right] P_{\vec{\theta'_i}}, \quad \forall a. $$
In this way, we can ``tilt'' the operator basis $\{\sigma_a\}$ at every interval, and still have enough freedom in $P_{\vec{\theta'}_i}$ to generate a non trivial switching function after the frame transformation. Similar to non-diagonal base sequences in~\cite{Paz2017}, this effectively grants us the ability to control which bath operator couples to which system operator.

Of particular relevance in the subsequent discussion will be the tilting transformation defined by $P_{\rm tilt} =[\frac{\pi}{4}]_x$. Under this transformation one finds that
\begin{align*}
\Big[ \frac{\pi}{4} \Big]_x :\sigma_{+} & \mapsto \sin^2 \frac{\pi}{8}\, \sigma_{-} +
i \, \frac{\sigma_0}{2}
+\cos^2 \frac{\pi}{8}\, \sigma_{+},\\
\sigma_0 & \mapsto \frac{1}{2} \Big(\!- i\, \sigma_{-}+
\sqrt{2} \, \sigma_0
+ i \, \sigma_{+} \Big),\\
\sigma_{-} & \mapsto \cos^2 \frac{\pi}{8}\, \sigma_{-} - i
\, \frac{\sigma_0}{2}
+\sin^2 \frac{\pi}{8}\, \sigma_{+},
\end{align*}
which in turn implies that in a given time interval where the transformation has been executed, all the switching functions $y_{j,j'} (t)$ are in general non-vanishing, i.e., they are non-diagonal as desired. While other tilting transformations are of course possible, the above is particularly convenient when the qubit quantization axis is along $\sigma_z$, as we will see in Sec.~\ref{Illus}.

\section{Protocols for comb-based multiaxis QNS}
\label{secProtocols}

\subsection{Accessible quantities}
\label{secAccessibleQuantities}

The first step to constructing an explicit QNS protocol is understanding what information is accessible to the experimentalist when the ability to prepare different initial states and measure in different bases is exploited and, in particular, which of the various integrals appearing in $\mathcal{C}^{(2)}_O$ can be isolated. For concreteness, we shall assume the ability to prepare any eigenstate of the Pauli basis, and measure in any direction. In other words, with $\rho_S \in \{ \eta_{\pm,\alpha} =({I_S \pm \sigma_{\alpha}})/{2}\}$, we assume we have access to all the expectation values 
$$ E(\sigma_\gamma (T))_ {\eta_{\pm,\alpha}\otimes \rho_B} =  {\rm tr}\Big[ e^{\mathcal{C}_\gamma}  \,
\frac{\sigma_{\gamma} \pm  \sigma_{\alpha} \sigma_\gamma }{2} \Big] ,$$
\noindent
where we have introduced the observable-dependent operator $\mathcal{C}_\gamma \equiv \sum_k (-i)^k {\mathcal{C}_{\sigma_\gamma}^{(k)}}/{k!}$ and $\alpha,\gamma \in \{x,y,z\}$. Now, by decomposing $\mathcal{C}_\gamma$ in the Pauli basis as 
\begin{equation}
\mathcal{C}_\gamma \equiv \sum_{\beta=0,x,y,z} C_{\gamma,\beta}  \; \sigma_\beta,
\label{expansion}
\end{equation}
one can see that by combining the different expectation values it is possible to isolate all $C_{\gamma,\beta}$ terms. For example, for $\sigma_{\gamma}= \sigma_x$, the experimentally accessible quantities $ \mathcal{M}^{(x)}_{r, \alpha} \equiv E(\sigma_x (T))_ {\eta_{+1,x}\otimes \rho_B} + r E(\sigma_x (T))_ {\eta_{-1,x}\otimes \rho_B}\ignore{\red{\langle \langle \sigma_x (T) \rangle \rangle \vert_{+,\alpha} + s \langle \langle \sigma_x (T) \rangle \rangle \vert_{-,\alpha}}}$, $r=\pm1$, are found to be given by 
\begin{align*}
\mathcal{M}^{(x)}_{+, x} &= \frac{2 e^{C_{x,0}} C_{x,x} \sinh \left(\sqrt{C_{x,x}^2+C_{x,y}^2+C_{x,z}^2}\right)}{\sqrt{C_{x,x}^2+C_{x,y}^2+C_{x,z}^2}},\\
\mathcal{M}^{(x)}_{-, x} &= 2 e^{C_{x,0}} \cosh \left(\sqrt{C_{x,x}^2+C_{x,y}^2+C_{x,z}^2}\right),\\
\mathcal{M}^{(x)}_{-, y}&= \frac{2 i e^{C_{x,0}} C_{x,z} \sinh \left(\sqrt{C_{x,x}^2+C_{x,y}^2+C_{x,z}^2}\right)}{\sqrt{C_{x,x}^2+C_{x,y}^2+C_{x,z}^2}},\\
\mathcal{M}^{(x)}_{-, z} &= -\frac{2 i e^{C_{x,0}} C_{x,y} \sinh \left(\sqrt{C_{x,x}^2+C_{x,y}^2+C_{x,z}^2}\right)}{\sqrt{C_{x,x}^2+C_{x,y}^2+C_{x,z}^2}}.
\end{align*} 
From these expressions, the quantities $C_{x,\beta}$ can be obtained by judiciously applying trigonometric identities, and a similar reasoning allows us to obtain the coefficients $C_{\gamma,\beta}$ for other observables. Recalling that we are truncating the cumulant expansion to order $2$, it follows from Eq. \eqref{expansion} that we then have access to all the quantities $$C_{\gamma,\beta} \simeq {\rm tr}[\mathcal{C}_{\sigma_\gamma}^{(2)} \sigma_\beta]/2.$$

\subsection{Balanced and imbalanced filters \label{secBalancedImbalanced}}

Given the access to the projections of $\mathcal{C}^{(2)}_{\sigma_\gamma}$ on each axis and Eqs.~\eqref{cumussphe}-\eqref{c2carte}, one recognizes that there are two classes of filters present in the problem. Namely, filters 
$$ G^{\pm}_{a,a';b,b'} (\omega + s \Omega, -\omega + s' \Omega, T),\quad s,s' \in \{-1,0,+1\}, $$
for which the frequency displacement induced by $H_S$ on the FF is {\it balanced}, when $(s+s') \Omega =0$, and for which it is {\it imbalanced}, when $(s + s') \Omega  \neq 0.$
This distinction is crucial. For a balanced displacement, $\omega + s \Omega = - (- \omega + s'\Omega)$ and thus one has filters of the form $G^{\pm}_{a,a';b,b'} (\omega', -\omega', T)$. For such filters, the discussion in Sec. \ref{basicspectro} suggests that one just needs to generate the necessary symmetries to produce a comb, i.e., if these were the only FFs entering the problem, then the results obtained for dephasing models in \cite{Paz2017} would in principle seamlessly extend to the general decoherence scenario. However, for imbalanced filters, which do not have the simple form $G^{\pm}_{a,a';b,b'} (\omega', -\omega', T),$ a comb {\em cannot} be generated by simply applying the aforementioned symmetries and a different treatment is needed.

\subsubsection{QNS and balanced filters}

As highlighted earlier, reconstructing the power spectra via a balanced filter requires the use of the aforementioned set of well-established symmetries and indeed, at a first glance, it would seem that it is necessary to be able to impose any of them. However, as it turns out, the fact that we are working with a single-qubit probe reduces the need for some of them.

The simplest scenario arises when $\Omega=0$, as in this case 
only $G^{\pm}_{a,a';b,b'} (\omega, -\omega, T)$ filters appear in the dynamics. Furthermore, a close analysis of Eq.~\eqref{c2carte}, shows that it is possible to combine observables in such way that {\em all} spectra are filtered by $G^{+}_{a,a';b,b'} (\omega, -\omega, T)$ filters. This is a striking difference between the two-qubit single-axis and the single-qubit multiaxis scenarios, which can be traced back to the the seemingly innocuous $f^\alpha_{\alpha'}$ superscript in Eq.~\eqref{c2carte}. For example, one can reconstruct ${\rm Re}[S_{x,z}^\pm(\omega )]$ and ${\rm Im}[S_{x,z}^\pm(\omega )]$ from the relations
\begin{align*} C_{x,y}-C_{z,y}&= 4 i \int_0^\infty \frac{ \mathrm{d}\omega}{2\pi}\Big({\rm Im}[S_{x,z}^+(\omega )] {\rm Im}[G_{x,x,z,z}^+(\omega ,T)]\\
& \,\,\,\,\,\, - {\rm Re}[S_{x,z}^+(\omega )] {\rm Re}[G_{x,x,z,z}^+(\omega ,T)] \Big),\\
C_{y,y}&=4\int_0^\infty \frac{ \mathrm{d}\omega}{2\pi} \Big({\rm Re}[S_{x,z}^-(\omega )] {\rm Im}[G_{x,x,z,z}^+(\omega ,T)]\\
& \,\,\,\,\,\, + {\rm Im}[S_{x,z}^{{-}}(\omega )] {\rm Re}[G_{x,x,z,z}^+(\omega ,T)]\Big),
\end{align*}
simply by using repetition symmetry. The remaining spectra can be obtained in a similar fashion by combining $C_{\alpha,\beta}$'s and using frame-tilting base sequences, i.e., both diagonal and non-diagonal control. We stress that this does not mean that $G^-$ filters do not enter the dynamics of the qubit (see Appendix~\ref{Vsplit}); they just are not necessary for QNS purposes.  

The general $\Omega\neq 0$ scenario shares some similar features. In particular, a detailed analysis of Eq.~\eqref{cumussphe} shows that, while both balanced and imbalanced filters are now present, one may still combine $C_{\alpha,\beta}$ quantities in a way that only $G^{+}_{a,a';b,b'} (\omega + s \Omega, -\omega + s'\Omega, T)$ filters are relevant to the QNS problem (we show this explicitly in the next section). In turn, this implies that, similar to the single-qubit single-axis problem~\cite{Alvarez2011,Norris2016}, sequence repetition suffices for QNS -- provided one can find a compatible mechanism to deal with the imbalanced $G^+$ filters, to which we turn now.

\subsubsection{QNS and imbalanced filters }
\label{imbal}

Imbalanced filters pose a different challenge. To see this, consider first the effect of repeating a given base sequence $M$ times. Then, the filters relevant to spectroscopy take the form
\begin{align*}
& G^+_{a,a';b,b'} (\omega + s \Omega, -\omega + s' \Omega, M T_c) = \\
& e^{i (s +s')\frac{\Omega T_c}{2}  }  \frac{\sin^2 (M \frac{\omega T_c}{2})}{\sin^2 (\frac{\omega T_c}{2})}  G^+_{a,a';b,b'} 
(\omega + s \Omega, -\omega + s' \Omega, T_c).
\end{align*}
The displacement in the frequency argument leads to an extra exponential factor that does {\em not} appear in the expression leading to the comb, Eq. \eqref{combg}. Formally, when $\Omega\neq 0$, we may envision imposing a {\it synchronization} condition between the qubit energy scale and the periodicity of the applied control to recover the desired comb-generating expression. That is, we may demand that
$ {\Omega T_c}/{2} = m \pi,$
for some integer $m$, such that the $e^{i (s+s')\frac{\Omega T_c}{2}} = \pm 1$. In other words, control repetition along with the synchronization condition are enough to generate a comb in all relevant filters for arbitrary values of $\Omega$.

\subsubsection{Imbalanced filters in the large qubit-splitting regime} 
\label{LargeQ}

While, mathematically, synchronization solves the problem, one needs to consider whether it can be realistically achieved. Even with perfect knowledge of $\Omega$, let us assume that the precision in timing is $\delta t$, that is, we can apply pulses at a times $t_i\pm \delta t$. In this situation, guaranteeing the synchronization condition implies then that ${\Omega \delta t}/{2} \ll \pi.$ In turn, this imposes the constraint $\delta t \ll {2 \pi}/{\Omega},$ which may be unrealistic when $\Omega$ is sufficiently large -- as is the case in many realistic devices (see also Sec. \ref{Illus}).

Fortunately, this potentially problematic regime has a redeeming property: as we now show, when $\Omega T$ is sufficiently large, {\em imbalanced filters become negligible}. Thus, spectra that enter the dynamics only via convolutions with such filters are effectively irrelevant. This result follows essentially from separating the two important timescales in the problem, namely, the evolution time $T$ and $1/\Omega$, similar in spirit to the secular approximation often used in open quantum systems~\cite{Breuer:book}. We provide the detailed mathematical derivation of this argument in Appendix~\ref{multiscale}.

The above implies that, crucially, when $\Omega T\gg 1$ and the imbalanced filters are negligible, the {\em only} spectra contributing to the qubit dynamics are $ S^\pm_{\pm,\mp} (\omega)$ and $S^\pm_{0,0}(\omega)$. Thus, these spectra are the only ones that need to be reconstructed in order to model the qubit dynamics. Interestingly, extending the argument presented here to higher order filters would imply that, in the $\Omega T \gg 1$ regime, the non-unitary effective propagator $\langle \mathcal{T} e^{-i \int_{-T}^T\widetilde{H}_O (s) ds } \rangle$ (see  Eqs.~\eqref{compact} and~\eqref{cumg}) acting on the qubit can always be written as $e^{A I_S + B \sigma_z}$. We highlight that despite this simple single-axis form, Eq.~\eqref{compact} allows for both dephasing and relaxation in the dynamics. 

\section{Illustrative results}
\label{Illus}

We are now ready to showcase the key mechanisms allowing us to perform QNS of a multiaxis noise model in a  concrete example. We will focus on the physically relevant large $\Omega$ regime. Our intention is three-fold: (i) show in detail the previous claims regarding the role of $G^+$-filters in multiaxis QNS, (ii) that only a limited set of spectra contribute to the dynamics in the $\Omega T \gg 1$ regime, and (iii) explicitly show numerical results that will allow us to discuss some of the potential limitations of the comb approach.

To start, we explore the structure of the accessible quantities $C_{\alpha,\beta}$. A direct calculation shows that the four combinations of experimentally accessible quantities $Q_1(T)=\frac{ C_{z,0} 
+ C_{z,z}}{2},Q_2(T)=\frac{ C_{z,0} 
- C_{z,z}}{2},Q_3(T)=C_{x,0}$ and $Q_4(T)=C_{x,x}$ are sufficient to reconstruct all the necessary spectra. This follows since
\begin{equation}
\label{Cexp}
Q_p(T) = \sum_{j,l=-1,0,1} \int_{-\infty}^\infty  \frac{ \mathrm{d} \omega}{2 \pi}\mathcal{G}^{(p)}_{j,l} (\omega,T)    S_{j,l}(\omega),\end{equation}
where the generalized FFs
$\mathcal{G}^{(\alpha,\beta)}_{j,l} (\omega,T) $ are given by 
\begin{widetext}
\begin{align*}
\mathcal{G}^{(1)}_{-j,-l}(\omega,T) &= G^{+}_{j,1;l,-1}(\omega + j \Omega, -\omega + l \Omega, T), \\ 
\mathcal{G}^{(2)}_{-j,-l}(\omega,T) &= G^{+}_{j,-1;l,1}(\omega + j \Omega, -\omega + l \Omega, T), \\ 
\mathcal{G}^{(3)}_{-j,-l}(\omega,T) &= 2 G^{+}_{-j,0;-l,0} (\omega + j \Omega, -\omega + l \Omega,T)- \frac{1}{2} \sum_{j',l' =-1,1} j' l'\,\,  G^{+}_{j,j';l,l'} (\omega + j \Omega, -\omega + l \Omega,T) , \\
\mathcal{G}^{(4)}_{-j,-l}(\omega,T) &= \sum_{\substack{j',l' =-1,0,1\\ |j'+l'| = 1}} (l'- j')\,\,  G^{+}_{j,j';l,l'} (\omega + j \Omega, -\omega + l \Omega,T).
\end{align*}
\end{widetext}
The form of the generalized filters above explicitly shows our previous claim: since all spectra are represented, then only $G^+$- type filters are necessary for QNS, and thus only sequence repetition is necessary to ensure the appearance of the frequency comb. Moreover, since we are concerned with the  $\Omega T \gg 1$ regime, terms involving imbalanced filters can be negelcted and the sum in Eq.~\eqref{Cexp} is effectively restricted by the condition $j + l=0$. 

\subsubsection{Filter design principles}
 Having established that indeed all of the integrals can be deconvolved via a frequency comb structure by the use of control repetition, the remaining key aspect of the protocol is that of filter design, i.e., finding control sequences whose filters are capable of sampling the full spectra. This is critically important in the case of quantum noise. 
 Note that the quantum components of the spectra, $S_{j,l}^-(\omega)$, and the classical components, $S_{j,l}^+(\omega)$, are odd and even functions of $\omega$, respectively. It follows from \erf{Cexp} that $S_{j,l}^-(\omega)$ can only be reconstructed if we design sequences such that the corresponding filter $\mathcal{G}^{(p)}_{j,l} (\omega,T)$ is odd in the frequency domain. 
 
In order to achieve this, we must understand the structure of the balanced $G^+$ filters. Consider that any function $f(\omega)$ can be written as 
$f(\omega) =  \mathcal{E} [f(\omega)]+\mathcal{O} [f(\omega)], $
where its even and odd components are given, respectively, by  
\begin{align*}
\mathcal{E} [f(\omega)] = \frac{f(\omega) + f(-\omega)}{2}, \quad 
\mathcal{O} [f(\omega)] = \frac{f(\omega) - f(-\omega)}{2}.
\end{align*}
By manipulating the integral expressions of the filters, we find
\begin{widetext}
\begin{align}
\label{eq::EG}
\mathcal{E}\big[G_{a,a';b,b'}^{+}(\omega,-\omega,T)\big]=&\int_{-T/2}^{T/2} \mathrm{d}s\,\int_{-T/2}^{T/2} \mathrm{d}s' \,
\Big\{\mathcal{E}\big[y_{a,a'}(s+T/2)\big]\mathcal{E}\big[y_{b,b'}(s'+T/2)\big]\cos(\omega s)\cos(\omega s')
\\&\quad\quad\quad\quad\quad\quad\quad\quad\quad\quad
+\mathcal{O}\big[y_{b,b'}(s'+T/2)\big]\mathcal{O}\big[y_{a,a'}(s+T/2)\big]\sin(\omega s')\sin(\omega s)\Big\}, 
\notag\\ \label{eq::OG}
\mathcal{O}\big[G_{a,a';b,b'}^{+}(\omega,-\omega,T)\big]=&i\int_{-T/2}^{T/2} \mathrm{d}s\,\int_{-T/2}^{T/2} \mathrm{d}s' \,
\Big\{\mathcal{O}\big[y_{a,a'}(s+T/2)\big]\mathcal{E}\big[y_{b,b'}(s'+T/2)\big]\sin(\omega s)\cos(\omega s')
\\&\quad\quad\quad\quad\quad\quad\quad\quad\quad\quad
-\mathcal{O}\big[y_{b,b'}(s'+T/2)\big]\mathcal{E}\big[y_{a,a'}(s+T/2)\big]\sin(\omega s')\cos(\omega s)\Big\}.
\notag
\end{align}
\end{widetext}
It is evident that the even/odd character of the FFs can be traced back to the even/odd character of the switching functions $y_{a,a'}(t + T/2)$ on the interval $[-T/2,T/2]$ or, equivalently, to the mirror symmetric/antisymmetric character of the switching functions in the interval $[0,T]$. If the switching functions $y_{a,a'}(s+T/2)$
and $y_{b,b'}(s'+T/2)$ have the same parity (either both even or both odd), then the odd component of the FF vanishes. Conversely, if $y_{a,a'}(s+T/2)$
and $y_{b,b'}(s'+T/2)$ have opposite parity, the even component of the FF vanishes.

\subsubsection{A sample reconstruction}
\label{secSampleReconstruction}

\begin{figure*}[t]
\centering
\includegraphics[width = 175mm]{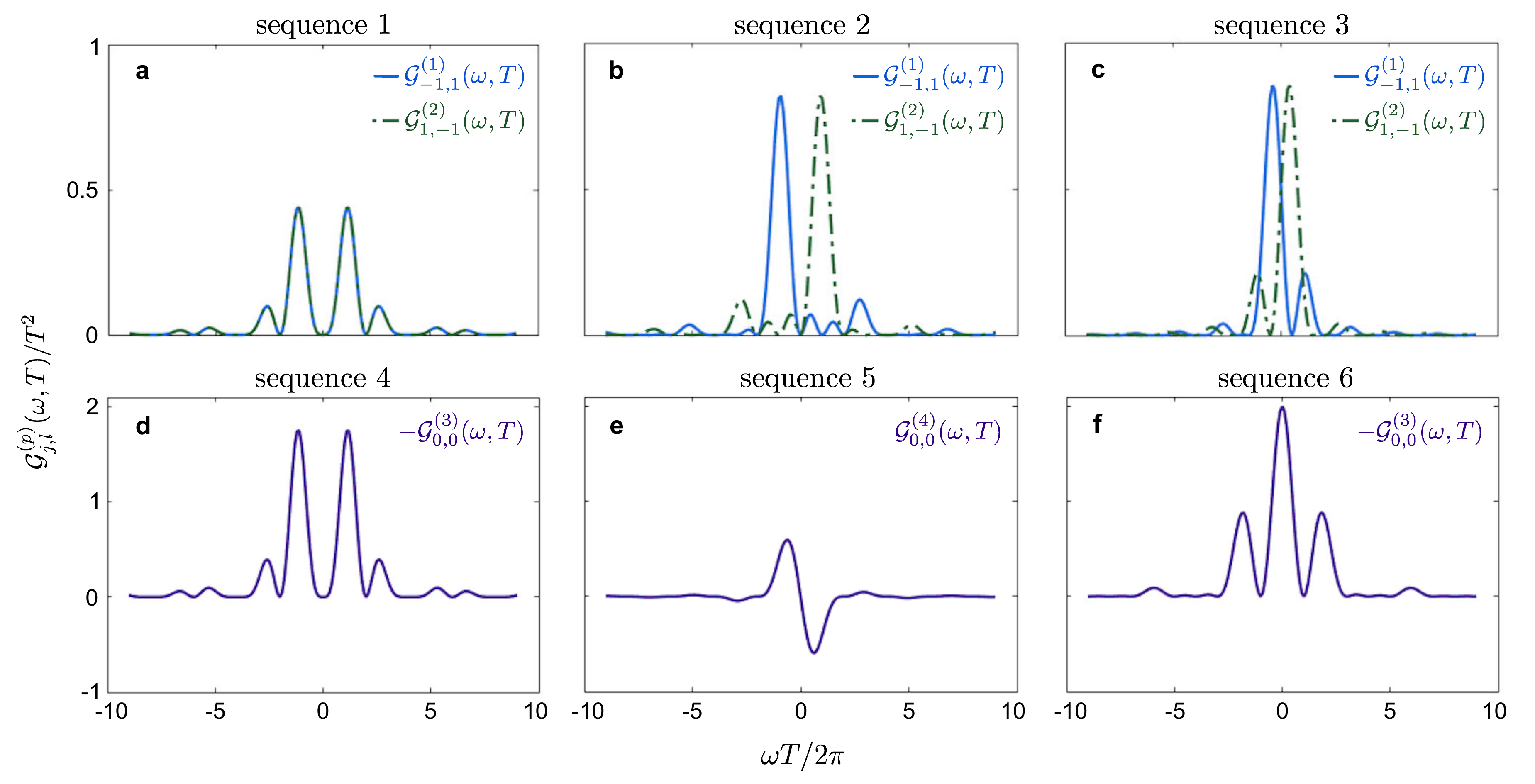}
\vspace*{-4mm}
\caption{(Color online) 
Top: FFs used to reconstruct the transverse noise spectra. The FFs coupling to $S_{-1,1}(\omega)$ in quantity 1 (top row) and $S_{1,-1}(\omega)$ in quantity 2 (bottom row) are plotted versus multiples of the harmonic frequency. Columns correspond to the diagonal control sequences used in the reconstruction with (a,d) sequence 1, (b,e) sequence 2, and (c,f) sequence 3. 
Bottom: FFs used to reconstruct the dephasing noise spectrum. FFs corresponding to (a) the diagonal sequence 4, (b) the non-diagonal sequence 5, and (c) the non-diagonal sequence 6 are plotted versus multiples of the harmonic frequency. The filters couple to $S_{0,0}(\omega)$ in quantities 3, 4, and 3, respectively.
}
\label{fig:FF1}
\end{figure*}

Finally, we combine all of the above tools and observations into a sample recipe for reconstructing the spectra influencing the qubit dynamics in the $\Omega T \gg 1$ regime. Concretely, we make use of the following six sequences over $[0,T_c]$ (in the interaction frame):
\begin{widetext}
\begin{align*}
    U_1
    & \!\equiv U_0\Big(\frac{3T_c}{4}, T_c\Big) \,
    [\pi]_z\, U_0\Big(\frac{T_c}{4}, \frac{3T_c}{4}\Big)\, 
    [\pi]_z\, U_0\Big(0, \frac{ T_c}{4}\Big), \\ 
    U_2
    & \!\equiv \Big[\frac{\pi}{2}\Big]_z\, U_0\Big(\frac{3 T_c}{4}, T_c\Big)\, 
    \Big[\frac{\pi}{2}\Big]_z\,
    U_0\Big(\frac{T_c}{2}, \frac{3 T_c}{4}\Big), 
    \Big[\frac{\pi}{2}\Big]_z\, U_0\Big(\frac{T_c}{4}, \frac{T_c}{2}\Big)\, 
    \Big[\frac{\pi}{2}\Big]_z\, 
    U_0\Big(0, \frac{ T_c}{4}\Big),\\
    U_3
    & \!\equiv  
    \Big[\frac{3\pi}{2}\Big]_z\, U_0\Big(\frac{T_c}{2}, T_c\Big) \,
    \Big[\frac{\pi}{2}\Big]_z\, U_0\Big({0,\frac{T_c}{2}}\Big)\\
    U_4
    & \!\equiv 
    [\pi]_z\,
    U_0\Big(\frac{3 T_c}{4}, T_c\Big) \, 
    [\pi]_y\, U_0\Big(\frac{ T_c}{2}, \frac{3 T_c}{4}\Big) \,
    [\pi]_z\, U_0\Big(\frac{ T_c}{4}\, \frac{ T_c}{2}\Big) \,
    [\pi]_y\, U_0\Big(0, \frac{ T_c}{4}\Big),\\
    U_5
    & \!\equiv  [\pi]_x\,\widetilde{U}_0\Big(\frac{ T_c}{2}, T_c\Big)\,[\pi]_{y}
    \widetilde{U}_0\Big(\frac{ T_c}{4}, \frac{T_c}{2}\Big) \,
    [\pi]_z\, \widetilde{U}_0\Big(0, \frac{ T_c}{4}\Big)\\
    U_6
    & \!\equiv 
    [\pi]_z\, \widetilde{U}_0\Big(\frac{3 T_c}{4}, T_c\Big) \,
    [\pi]_z\, \widetilde{U}_0\Big(\frac{T_c}{2}, \frac{3T_c}{4}\Big) \,
    [\pi]_z \,\widetilde{U}_0\Big(\frac{ T_c}{4}, \frac{T_c}{2}\Big) \,
    [\pi]_z\, \widetilde{U}_0\Big(0, \frac{ T_c}{4}\Big),
    \end{align*}
\end{widetext}
where $\widetilde{U}_0(t_i,t_j) = 
[-{\pi}/{4}]_x {U}_0(t_i,t_j) 
[{\pi}/{4}]_x$ denotes free evolution in the interaction picture tilted by $[{\pi}/{4}]_x$ The first three sequences will be used to reconstruct the $S_{\pm, \mp} (\omega)$ spectra in tandem with the accessible quantities $Q_1$ and $Q_2$, while the remaining ones will be used with $Q_3$ and $Q_4$ in the reconstruction of $S_{0,0}(\omega)$. As can be seen from Figs.~\ref{fig:FF1}, the chosen sequences are such that we have access to odd and even filters. This and the fact that the frame tilting sequences, $U_5(t)$ and $U_6(t)$, generate non-trivial non-diagonal switching functions, provide the necessary tools to fully reconstruct the spectra as we demonstrate in our numerics.

\begin{figure}
\hspace*{-.4cm}
\includegraphics[width = 90mm]{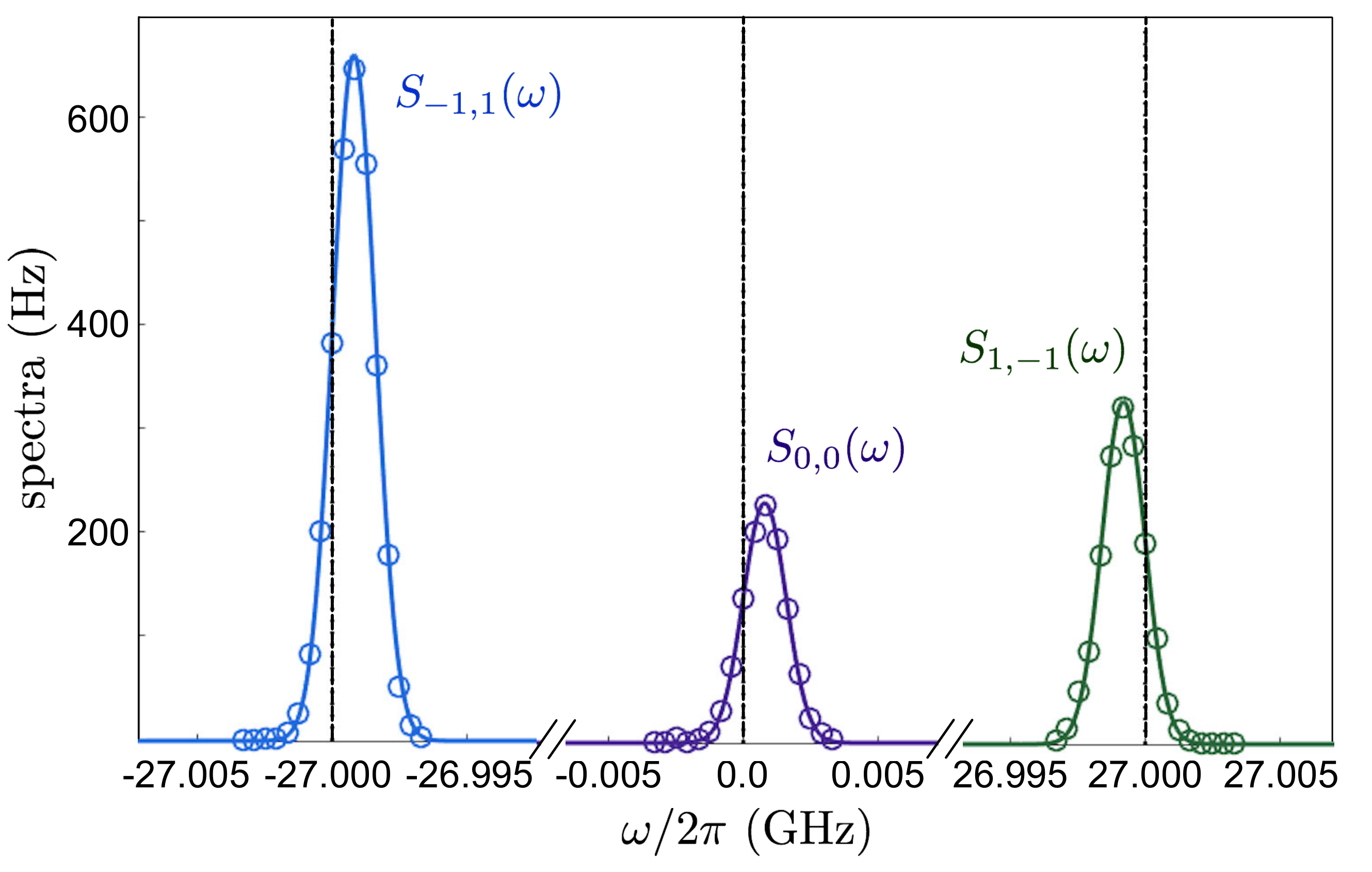}
\vspace{-.5cm}
\caption{(Color online) Spectral reconstructions. From left to right, the transverse noise spectrum $S_{-1,1}(\omega)$ (blue solid line) and reconstruction $\hat{S}_{-1,1}(\omega)$ (blue circles) are plotted in a frequency band centered at $\omega=-\Omega$, the dephasing noise spectrum $S_{0,0}(\omega)$ (purple solid line) and reconstruction $\hat{S}_{0,0}(\omega)$ (purple circles) are plotted in a frequency band centered at $\omega=0$, and the transverse noise spectrum $S_{1,-1}(\omega)$ (green solid line) and reconstruction $\hat{S}_{1,-1}(\omega)$ (green circles) are plotted in a frequency band centered at $\omega=\Omega$. Dashed vertical lines, from left to right, correspond to the frequencies $\omega=-\Omega$, $\omega=0$, and $\omega=\Omega$.  }
\label{fig:Rec}
\end{figure}

Following the recipe for the frequency comb approach detailed in Sec.~\ref{secTools} each of the  base sequences above will be applied periodically during the spectroscopy procedure. Concretely each experiment consists of $M=20$ repetitions of the basic sequence $U_i$, for different cycle times $T_{c}=2.4,\,2.4/2,\ldots,2.4/8\,\mu$s. We use these sequences to probe the noise coupled to a qubit with energy splitting $\Omega/2\pi=27$ GHz (notice that $\Omega MT_c \gg 1$ as desired). In the Cartesian representation, the spectrum of the noise is a sum of Gaussians centered at different locations in the frequency domain, 
\begin{align*}
S_{\alpha,\beta}(\omega) &= Ae^{-\frac{(\omega-\omega_-)^2}{2\Delta^2}}+0.7Ae^{-\frac{(\omega-\omega_0)^2}{2\Delta^2}}+0.5Ae^{-\frac{(\omega-\omega_+)^2}{2\Delta^2}},
\end{align*}
where $\alpha,\beta\in\{x,y,z\}$, $A=332$ Hz, $\Delta/2\pi = 0.80$ MHz, $(\omega_\mp \pm\Omega) /2\pi = \pm 0.81$ MHz and $\omega_0/2\pi=0.80$ MHz. We are ultimately interested in reconstructing the spherical spectra, which simplify to $S_{\pm,\mp}(\omega)=S_{x,x}(\omega)+S_{y,y}(\omega)$
since the $S_{\alpha,\beta}(\omega)$ are purely real. The number of different cycle times per base sequence naturally restricts the reconstruction window in comb-based protocols, as discussed in Refs.~\cite{Norris2016,Paz2017}. Here, with 8 different cycle times per base sequence, we can sample
$S_{\pm 1,\mp 1}(\omega)$ in the frequency windows $[-\Lambda\pm\Omega,\Lambda\pm\Omega]$ and $S_{0,0}(\omega)$ in the frequency window $[-\Lambda,\Lambda]$, where $\Lambda=8\omega_0$ spans the width of 8 harmonics. Figure ~\ref{fig:Rec} shows plots of the actual and reconstructed spherical spectra in their respective frequency windows. Note that $S_{-1,1}(\omega)$, $S_{0,0}(\omega)$ and $S_{1,-1}(\omega)$ are asymmetric about $\omega=-\Omega$, $\omega=0$ and $\omega=\Omega$, respectively, a signature of quantum noise.
The remarkable accuracy of the sampling supports our formal claim in ~\ref{imbal} regarding the negligible contribution of certain spectra in the $\Omega T \gg 1$ regime. Notice that the expectation values we have used, i.e., our simulated  experimental values, have been calculated using the contribution of all spectra, and thus if the terms we assumed to be negligible in the reconstruction were not so, we would have significant deviations in our sampling.   

\section{Limitations of comb-based QNS}  
\label{secLimitations}

Finally, we conclude our paper by discussing one of the limitations of the protocol. We do not intend this as a critique to the protocol itself, but rather, in the same spirit of recent work~\cite{szankowski2018accuracy}, as a way to highlight the regimes in which it is applicable. We consider this to be crucial, as applying the protocol blindly, i.e., without regard of the validity of the assumptions behind it, may lead to misleading spectral reconstructions, as we demonstrate below. 

As explained in Sec.~\ref{basicspectro}, a key step in QNS protocols is ensuring the convolutions involving the filters and the spectra turn into discrete and, importantly, truncated sums (as in Eq.~\ref{discrete}). With this, one can then build a linear system of equations from which the spectral information is extracted. In comb-based QNS protocols, this is achieved via the introduction of certain control symmetries, mainly repetition as was the case along this paper. Crucially, the comb effectively generates a ``sampling grid'' in frequency with points at multiples of a sampling frequency $\omega_0  =2 \pi/T_c$, which is fundamentally upper-bounded by the minimum time resolution in any experiment, and lower-bounded by the maximum possible evolution time which is necessarily finite given the presence of decoherence mechanisms. Importantly, this implies that there is a physical upper bound on the sampling frequency. 

\begin{figure}[t!]
\centering
\includegraphics[width = 75mm]{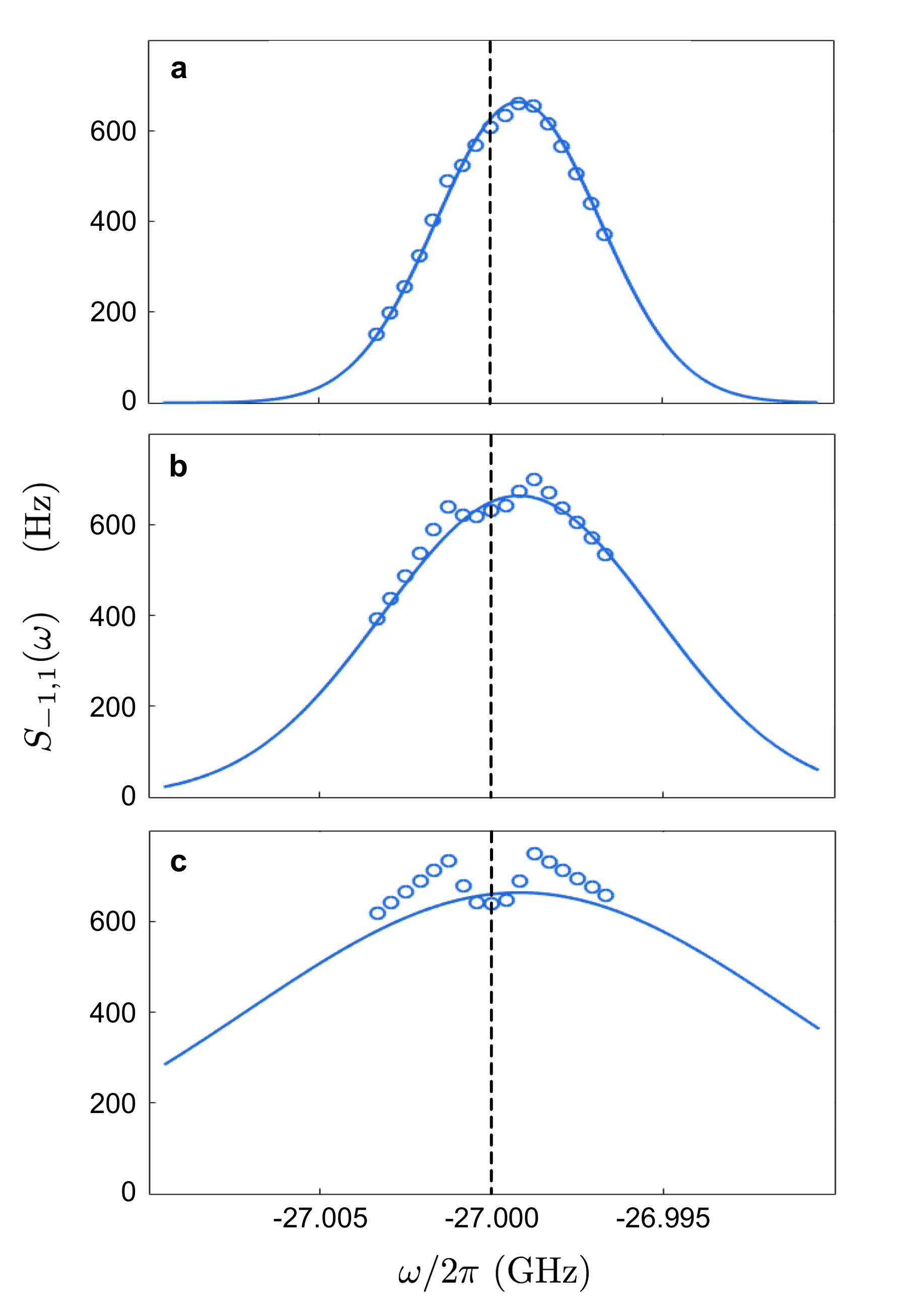}
\vspace*{-.2cm}
\caption{(Color online) Effect of increased bandwidth. In a frequency band centered at $\omega=-\Omega$, three different transverse noise spectra (blue solid lines) are plotted along with their corresponding reconstructions (blue circles). The transverse spectra are Gaussian of varying bandwidth with (a) $\Delta/2\pi=2.4$ MHz, (b) $\Delta/2\pi=4.0$ MHz, and (c) $\Delta/2\pi=8.0$ MHz. }
\label{fig:MoreBandwidth}
\end{figure}

This can lead to problems when one wants to sample a spectrum that has a very wide support in frequency space, as we now explain. Imagine that the spectrum has support in $\omega \in [0,\Omega_{\rm cutoff}]$. Now, the first step in a successful discretization is guaranteeing that the convolution is well approximated by the discrete sum. This entails, as Eq.~\eqref{discrete} suggests, that one has to guarantee that $$K \omega_0 \geq \Omega_{\rm cutoff},$$ 
that is, that the reconstruction window is large enough to sample the full support/bandwidth of the spectrum. Clearly, this implies that given $\Omega_{\rm cutoff}$ and the upper bound on $\omega_0$ implied by the physical constraints on the control, there is a lower bound, $K_0$, to the value of $K$, and thus on the size of our linear system. 

The problem arises then when $K_0$ is very large. This implies that one has to find at least $K_0$ sufficiently ``different'' control sequences that ensure that the resulting linear system is well-conditioned and allows for a faithful and robust recovery of the spectral information. Since the FFs themselves decay with frequency, even though one may in principle displace the center of the filter in frequency, it becomes challenging to build a well-conditioned system of equations. This limitation is one that has to be taken into account when designing the spectroscopy protocol. 

We stress that this is not an issue that is exclusive to multiaxis noise and, indeed, we tangentially discussed this question in a previous work~\cite{Norris2016}, which considered a pure-dephasing regime. However, in practical applications, and in particular in the context of solid-state devices, dephasing noise is typically associated with spectra that are strongly concentrated at low frequencies, which has proven to be favorable to comb-based QNS techniques. In both superconducting qubits and spin qubits, on the contrary, spectroscopy of dephasing noise has revealed spectra that are strongly suppressed with frequency, typically following power laws $\sim 1/\omega^{0.8\mbox{-}2.5}$ arising from charge noise~\cite{yoneda2018quantum,chan2018assessment}, flux noise~\cite{Bylander2011,yan2016flux}, nuclear spins~\cite{malinowski2017spectrum}, or external magnetic field instabilities~\cite{Muhonen2014}. By contrast, in these systems, noise sources leading to qubit relaxation (here corresponding to noise along $x$, $y$, or both) are often associated with spectra that \emph{grow} with frequency, up to a cutoff that can be orders of magnitude above the qubit splitting. For example, for spin qubits in semiconductors, phononic environments play a prominent role in qubit relaxation~\cite{meunier2007experimental,yang2013spin,petit2018spin}, leading to spectra whose frequency dependencies correspond to rapidly increasing power laws~\cite{tahan2014relaxation}. Qubit-relaxation phenomena in the solid state also frequently involve Ohmic noise spectra ($\propto\omega$). Examples of processes that can be associated with Ohmic spectra include Johnson-Nyquist noise afflicting spin qubits~\cite{huang2014electron,petit2018spin}, and two-level fluctuators coupled to superconducting qubits, for which a crossover between $1/f$ and Ohmic noise is typically observed in the GHz frequency range~\cite{shnirman2005low,Quintana2017,yan2016flux,Bylander2011}. Comb-based QNS approaches are particularly ill-suited for such noise spectra; indeed, those techniques would then require including harmonics over a frequency support that can well be hundreds of GHz or more (up to the cutoff frequency), while control is rarely faster than 10~ns, leading to harmonics separated by only $\sim 100$ MHz. This would imply designing well-conditioned reconstructions including several thousands of pulse sequences. Therefore, we argue that the shortcomings of the comb approach described above will manifest themselves most strongly in a multiaxis setting, in which dephasing and relaxation noise are simultaneously characterized.

We showcase this effect by fixing the size of the reconstruction window and attempting to reconstruct spectra with increasingly large support in frequency space. The results, depicted in Fig.~\ref{fig:MoreBandwidth}, show that as the support of the spectrum becomes larger the quality of the reconstruction decreases. More importantly, they support our previous warning: the choice of control sequences is dictated by the assumed width of the spectrum and this is an important consideration that has to be taken into account when implementing comb-based QNS protocols. These limitations can be in principle overcome even in the general decoherence scenario by switching to a more flexible continuous-time control paradigm (see for example Ref. ~\cite{willick2018efficient} for a continuous-drive dephasing-only spectroscopy protocol), and this is one of the objectives of our current efforts~\cite{MultiAxisNext}.

\section{Conclusion}

We have presented a DD comb-based spectroscopy protocol capable of characterizing the noise affecting a qubit in all directions, i.e., a qubit undergoing general decoherence. This significantly increases our ability to understand, and eventually control, the different types of noise processes leading to corruption of information in a qubit, as we are now capable of simultaneously reconstructing the power spectra of noise inducing $T_1$ and $T_2$ related processes, including their possible cross-correlations. 

In particular, we showed how to extend  comb-based noise spectroscopy methods to the general decoherence scenario for a series of regimes of increasing complexity. The essence of our result is recognizing that in each regime {\it all relevant} power spectra can be accurately reconstructed. Moreover, we laid out some control design fundamental principles allowing the reconstruction of the various power spectra associated to general noise models, thus considerably generalizing our previous results applicable to dephasing-only models. 

In developing our protocol, we have purposefully restricted ourselves to control symmetries compatible with our previous results on spectroscopy of multiqubit and non-Gaussian dephasing models~\cite{Paz2017,Norris2016}. In principle, this should allow us to merge them with the result presented here in a straightforward manner, thus achieving the coveted general multiaxis and multiqubit spectroscopy. This is part of our current efforts. On the experimental side, we expect that our results will open new avenues for high quality control of quantum systems, as they give access to all the noise information relevant to the dynamics of a qubit. Indeed, similar experiments to the ones performed in platforms where dephasing noise is dominant~\cite{chan2018assessment, Muhonen2014,sung2019non}, should now also possible in platforms where both $T_1$ and $T_2$ processes are significant. 

\section*{Acknowledgements}

This research was funded by the U.S. Army Research Office grant No. W911NF-141-0682, the U.S. MURI grant No. W911NF1810218 (to L.V.), the AUSMURI grant No. AUSMURI000002 and the DECRA fellowship DE170100088 (to G.P.S.). F.B. also gratefully acknowledges support from the {\em Fonds de Recherche du Qu{\'e}bec-Nature et Technologies}. 

\appendix
\begin{widetext}
\section{ Accessible quantities for vanishing qubit splitting}
\label{Vsplit}

In the main text we showed that only $G^+$ filters are necessary for QNS protocols. This does not mean, however, that $G^-$ filters do not contribute to the reduced dynamics. Indeed, an explicit calculation shows that, in general, both types of filters can contribute. We find:
\begin{align*}
C_{x,0}&= -2 \int_0^\infty \frac{ \mathrm{d}\omega}{2\pi} ({\rm Re}[S_{y,y}^+(\omega )] {\rm Re}[G_{y,y,y,y}^+(\omega ,T)]+{\rm Re}[S_{z,z}^+(\omega )] {\rm Re}[G_{z,z,z,z}^+(\omega ,T)]),\\
C_{y,0}&=-2 \int_0^\infty \frac{ \mathrm{d}\omega}{2\pi}({\rm Re}[S_{x,x}^+(\omega )] {\rm Re}[G_{x,x,x,x}^+(\omega ,T)]+{\rm Re}[S_{z,z}^+(\omega )] {\rm Re}[G_{z,z,z,z}^+(\omega ,T)]),\\
C_{z,0}&=-2 \int_0^\infty \frac{ \mathrm{d}\omega}{2\pi}({\rm Re}[S_{x,x}^+(\omega)] {\rm Re}[G_{x,x,x,x}^+(\omega ,T)]+{\rm Re}[S_{2,2}^+(\omega )] {\rm Re}[G_{2,2,2,2}^+(\omega ,T)]),\\
C_{x,y}-C_{z,y}&= 4 i \int_0^\infty \frac{ \mathrm{d}\omega}{2\pi}({\rm Im}[S_{x,z}^+(\omega )] {\rm Im}[G_{x,x,z,z}^+(\omega ,T)]-{\rm Re}[S_{x,z}^+(\omega )] {\rm Re}[G_{x,x,z,z}^+(\omega ,T)]),\\
C_{x,z}-C_{y,z}&= -4 i \int_0^\infty \frac{ \mathrm{d}\omega}{2\pi}({\rm Im}[S_{x,y}^+(\omega )] {\rm Im}[G_{x,x,y,y}^+(\omega ,T)]-{\rm Re}[S_{x,y}^+(\omega )] {\rm Re}[G_{x,x,y,y}^+(\omega ,T)]),\\
C_{z,x}-C_{y,x}&= 4 i \int_0^\infty \frac{ \mathrm{d}\omega}{2\pi}({\rm Im}[S_{y,z}^+(\omega )] {\rm Im}[G_{y,y,z,z}^+(\omega ,T)]-{\rm Re}[S_{y,z}^+(\omega )] {\rm Re}[G_{y,y,z,z}^+(\omega ,T)]),\\
C_{x,y}+C_{z,y}&= -4 i \int_0^\infty \frac{ \mathrm{d}\omega}{2\pi}({\rm Im}[S_{x,z}^+(\omega )] {\rm Im}[G_{x,x,z,z}^-(\omega ,T)]-{\rm Re}[S_{x,z}^+(\omega )] {\rm Re}[G_{x,x,z,z}^-(\omega ,T)]),\\
C_{x,z}+C_{y,z}&= 4 i \int_0^\infty \frac{ \mathrm{d}\omega}{2\pi}({\rm Im}[S_{x,y}^+(\omega )] {\rm Im}[G_{x,x,y,y}^-(\omega ,T)]-{\rm Re}[S_{x,y}^+(\omega )] {\rm Re}[G_{x,x,y,y}^-(\omega ,T)]),\\
C_{z,x}+C_{y,x}&= 4 i \int_0^\infty \frac{ \mathrm{d}\omega}{2\pi}({\rm Im}[S_{y,z}^+(\omega )] {\rm Im}[G_{y,y,z,z}^-(\omega ,T)]-{\rm Re}[S_{y,z}^+(\omega )] {\rm Re}[G_{y,y,z,z}^-(\omega ,T)]),\\
C_{x,x}&=4 \int_0^\infty \frac{ \mathrm{d}\omega}{2\pi}({\rm Re}[S_{y,z}^-(\omega )] {\rm Im}[G_{y,y,z,z}^+(\omega ,T)]+ {\rm Im}[S_{y,z}^{{-}}(\omega )] {\rm Re}[G_{y,y,z,z}^+(\omega ,T)]),\\
C_{y,y}&=4\int_0^\infty \frac{ \mathrm{d}\omega}{2\pi} ({\rm Re}[S_{x,z}^-(\omega )] {\rm Im}[G_{x,x,z,z}^+(\omega ,T)]+ {\rm Im}[S_{x,z}^{{-}}(\omega )] {\rm Re}[G_{x,x,z,z}^+(\omega ,T)]),\\
C_{z,z}&=4\int_0^\infty \frac{ \mathrm{d}\omega}{2\pi} ({\rm Re}[S_{x,y}^-(\omega )] {\rm Im}[G_{x,x,y,y}^+(\omega ,T)]+ {\rm Im}[S_{x,y}^{{-}}(\omega )] {\rm Re}[G_{x,x,y,y}^+(\omega ,T)]).
\end{align*}
As one can see from the above equations, the expectation value of any given observable, given an arbitrary initial state, manifestly depends on both $G^+$ and $G^-$ filters. What makes the multiaxis dynamics special is that there are observable-initial state combinations that depend solely on $G^+$ filters, which is in stark contrast with a multiqubit scenario under pure dephasing \cite{Paz2017}
. 

\section{On the vanishing of imbalanced filters in the large $\Omega T $ regime.}
\label{multiscale}

In the text, we provided the intuition behind the observation that when the splitting $\Omega$ is large in an appropriate sense, then certain filters and the corresponding spectra do not contribute to the probe dynamics. Here we show this formally.

To see the argument in detail, note that the filters $G^\pm$ are linear combinations of the second-order filters $F^{(2)}_{a,a';b,b'} (\omega + s \Omega, -\omega + s' \Omega, T)$. The filter is balanced when $s+s'=0$ and imbalanced otherwise (whenever $s+s' =\pm1$ or $s+s'=\pm2$). By redefining the integration variables in terms of $t_\pm \equiv (t_1 \pm t_2)/2$, 
it follows then that we can rewrite the filter as
\begin{align}\label{eq::FilterExpression}
& F^{(2)}_{a,a';b,b'} (\omega + s \Omega, -\omega + s' \Omega, T)/2 = \int_{\Lambda_{\pm}} \!\!\!\!\!\!  \mathrm{d}\vec{t}\, e^{i (2 \omega + (s- s') \Omega)  t_- + i (s+s') \Omega t_+} y_{a,a'}(t_+\!\! +\! t_-) y_{b,b'}(t_+\!\! - \!t_-),
\end{align}
where $\Lambda_{\pm}$ is the triangular integration domain defined by the vertices  $\{(t_-,t_+)\}=\{(0,T),(T/2,T/2),(0,0)\}$ or, alternatively, by the relations $t_{-} \in [0,T/2]$ and $t_+ \in [t_-, T-t_-]$. Let us further assume that the applied control induced switching functions are ``slow'' compared to $\Omega$. That is to say, $y_{c,c'}(t_+\pm t_-)$ for $c=a, b$ is well approximated by its truncated inverse Fourier transform, e.g.,
\begin{align*}
y_{c,c'}(t_+\pm t_-) \simeq \int_{-\Omega_0}^{\Omega_0} \frac{\mathrm{d}\omega^{(\vec{c})}_+}{2\pi}e^{i\omega^{(\vec{c})}_+t_+}\int_{-\Omega_0}^{\Omega_0} \frac{\mathrm{d}\omega^{(\vec{c})}_-}{2\pi}e^{i\omega^{(\vec{c})}_-t_-}\, \hat{y}_{c,c'}^\pm(\omega^{(\vec{c})}_+,\omega^{(\vec{c})}_-),
\end{align*}
where $\hat{y}_{c,c'}^\pm(\omega^{(\vec{c})}_+,\omega^{(\vec{c})}_-)$ represents the Fourier transform of $y_{c,c'}(t_+\pm t_-)$ and $\Omega_0 \ll \Omega$ is the bound of integration. By using  this expression to rewrite $y_{a,a'}(t_++ t_-)$ and $y_{b,b'}(t_+- t_-)$,
 the FF in \erf{eq::FilterExpression} becomes 
\begin{align*}
F^{(2)}_{a,a';b,b'} &(\omega + s \Omega, -\omega + s' \Omega, T)/2 = \mathcal{I}(o_+,o_-) \;  \hat{y}_{a,a'}^+(\omega^{(\vec{a})}_+,\omega^{(\vec{a})}_-)\, \hat{y}_{b,b'}^-(\omega^{(\vec{b})}_+,\omega^{(\vec{b})}_-)
\end{align*}
where 
\begin{align}
&\mathcal{I}(o_+,o_-)\equiv \int_{\Lambda_{\pm}} \!\!  \mathrm{d}\vec{t}\; e^{i t_+ o_+}\;e^{i t_- o_-}\notag,\\
&o_-\equiv  2 \omega + (s- s')\Omega - (\omega^{(\vec{a})}_- + \omega^{(\vec{b})}_-),\notag
\\
&o_+\equiv (s+s')\,\Omega - (\omega^{(\vec{a})}_+ + \omega^{(\vec{b})}_+).
\label{eq::oplus}
\end{align}
By letting $\alpha=o_+/o_-$, substituting $o_- = (1/\alpha) o_+$ into $\mathcal{I}(o_+,o_-)$,  and performing the time integration, we find 
$$ |\mathcal{I}(o_+,o_-)|\!\leq \!
\begin{cases} 
\!\frac{T^2}{(o_+ T)^2}\Big(\vert \frac{\alpha }{1+\alpha}\vert  \!+\! \vert \frac{ 2 \alpha^2}{-1+\alpha^2} \vert \!+\! \vert \frac{ \alpha }{-1+\alpha}\vert\Big),&\!\! \alpha \neq 1, \\
\!T^2 \Big(\frac{1}{(o_+ T)^2} +\vert \frac{1}{2 o_+ T}\vert \Big), &\!\! \alpha =1.
\end{cases}$$
From this expression, it is apparent that $|\mathcal{I}(o_+,o_-)|$ and, hence,
$|F^{(2)}_{a,a';b,b'} (\omega + s \Omega, -\omega + s' \Omega, T)|$ are small whenever $|o_+ T| \gg 1$. 

To verify that this condition is met, recall that $|\omega^{(\vec{a})}_+|, |\omega^{(\vec{b})}_+|\leq\Omega_0\ll\Omega$ due to our assumption of slow control. From \erf{eq::oplus}, it then follows that
$|o_+|\geq|\,|s+s'|\, \Omega - 2 \Omega_0|$. This translates into the observation that whenever
\begin{align}
    T|o_+|\geq\; T\Big|\;|s+s'|\, \Omega - 2 \Omega_0\Big| \gg 1,\notag 
\end{align}
the contributions imbalanced filters are negligible compared to those of balanced filters, for both $G^+$ and $G^-$. Since $\Omega\gg\Omega_0$ by the assumption of slow control, we can practically neglect the imbalanced filters whenever $\Omega T\gg 1$.
\end{widetext}

\bibliography{MultiaxisComb}

\end{document}